\documentclass[11pt]{article}

\usepackage{amsfonts,amsmath,amssymb,graphics}

\newcommand{\figscal}{0.9}

\newcommand{\C}{\mathbb{C}}

\newcommand{\I}{\mathrm{i}}
\newcommand{\pr}{\mathbf{P}}
\newcommand{\ex}{\mathbf{E}}

\newcommand{\notthis}[1]{}

\newcommand{\inv}{^{-1}}

\newcommand{\half}{\frac{1}{2}}

\newcommand{\shalf}{{\textstyle\frac{1}{2}}}

\newcommand{\mom}[1]{\langle #1 \rangle}
\newcommand{\bigmom}[1]{\left< #1 \right>}
\newcommand{\pd}{M}
\newcommand{\tr}{\mathrm{tr}}
\newcommand{\sig}{\varphi}
\newcommand{\bz}{\mathbf{z}}
\newcommand{\s}{;$\,$}

\newcommand{\demarcate}{\!\!\!\!\!\!\begin{array}{l}\left.\begin{array}{c} \\ \\ \\ \end{array}\right\} \mbox{$\pd$ rows} \\ \\ \begin{array}{c} \\ \\ \\ \\ \end{array}   \end{array}}

\begin{document}

\title{\bf Design and analysis of momentum trading strategies}
\author{Richard J. Martin\footnote{Dept.\ of Mathematics, Imperial College London, South Kensington, London SW1 2AZ, UK. Email: {\tt richard.martin1@imperial.ac.uk}}}
\maketitle

\begin{abstract}

We give a complete description of the third-moment (skewness) characteristics of both linear and nonlinear momentum trading strategies, the latter being understood as transformations of a normalised moving-average filter (EMA). We explain in detail why the skewness is generally positive and has a term structure.

This paper is a synthesis of two papers published by the author in RISK in 2012, with some updates and comments.

\end{abstract}


\section*{Introduction}

Trend-following, or momentum, strategies have the attractive property of generating trading returns with a positively skewed statistical distribution. Consequently, they tend to hold on to their profits and are unlikely to have severe `drawdowns'. They are very scalable and are employed in most asset classes---most traditionally in futures, where they are a favourite strategy among CTA (`commodity trading advisor') firms, but also in OTC markets---and by both buy- and sell-side practitioners.

The basic premise behind momentum is to buy what has been going up recently, and sell what has been going down. In other words, if recent returns have been positive then future ones are more likely to be positive, and similarly with negative. Systematic strategies formalise this notion by (i) measuring momentum, essentially by smoothing out recent returns to obtain a signal that is not too rapidly-varying, and (ii) having a law that turns this signal into a trading position, i.e.\ how may contracts or what notional to have on. Put this way, the ideas that only the finest minds can understand CTA strategies, or how the theory of statistics is of central importance to their construction, or how one needs to have been steeped in managed futures for many years to build a workable strategy, are seen to be self-serving and pretentious---a conclusion implicitly arrived at, even if not thus expressed, by other authors.

Before talking about skewness we may as well deal with the first moment, that is to say the expected return. it is important to understand that this is an entirely separate matter. Any statement about this depends on markets exhibiting momentum, i.e.\ serially-correlated returns. This can be ascribed to the way information is disseminated into markets or of behavioural characteristics of market participants. 
However, it is entirely subjective and is a matter of believing that markets will continue to behave in the way that they have done in the past. In contrast, as we analyse in detail here, even if market returns exhibit no serial correlation, during which period the strategy will produce no average return, the \emph{trading} returns of a momentum strategy will still have positive third moment. What is interesting is that the skewness characteristic is a product of the design of the strategy, whereas in long-only equities and credit the (negative) skewness is an intrinsic feature of the asset class that has to be tolerated and risk-managed.

Positive skewness results from the way positions are taken. Suppose that we look at the trading returns from one particular instrument (perhaps US Tsy bond futures) of a particular period (perhaps one week) over a long period of history. Let us group these returns by the size of the underlying position. The magnitude of P\&L will typically be larger when the magnitude of the position is larger, but crucially it will typically be positive too. This is because momentum strategies typically run bigger positions \emph{when they have already made money}---as opposed to reversion strategies that follow the opposite principle.
In statistical language the full distribution of P\&L is a mixture of distributions of different mean and variance: the components with a higher variance have positive mean and that is a recipe for positive skewness.

Studies on the subject have generally been empirical (for a good overview see e.g.\ \cite{Till11} and references therein, and \cite{AcarSatchell02} for a general introduction to technical trading).
However, there is a decent literature on quantitative aspects.
The first work was by Acar \cite{Acar92} who derived a variety of results in discrete time using different forecasting models and also ascertained that the distribution of momentum trading returns has positive skewness. Potters \& Bouchaud \cite{Potters05} consider a particular type of momentum strategy and derive rigorous results about its performance. 
Bruder et~al.\ \cite{Bruder11}, and an even longer extension by Jusselin et~al.\ \cite{Jusselin17}, devote considerable effort to deriving moving-average filters from an underlying model. In practice, however, this seems to create more problems than it solves because the assumed model may be wrong: better, we think, is to use a convenient definition of moving-average filter, here the exponentially-weighted moving average (EMA) as it is easily calculated by recursion, and design a strategy using those.
Then in 2012 two papers were published in RISK by the author, which considerably broadened the scope of the subject. Both focused on the third-moment characteristics of momentum models. The first \cite{Martin12b} dealt with linear models, by which we mean a signal proportional to a momentum signal obtained by applying exponential smoothing to the market returns. The second \cite{Martin12c} showed how to deal with strategies defined as nonlinear transformations of momentum signals, within the same framework. This paper is a synthesis of these two. More recently Dao \cite{Dao16} focuses on the connection between convexity, option-like characteristics, and momentum strategies.

In building momentum strategies, two main considerations are important. The first relates to backtesting, in other words finding what worked best in the past and assuming that it will continue to do so.  Part of the problem with this is that it is too reliant on historical data: if left unchecked, it wastes an inordinate amount of time in fitting and overfitting, mainly because different models typically produce almost identical historical performance.
The second relates to design: that is to say, without regard to the past, force the strategy to have certain statistical properties when the market behaves in predefined ways. One idea, which we consider in depth, is when markets are not trending (and so market returns are uncorrelated). Another is when the market does trend in a way that it did in a particular historical scenario, such as gold in the first decade of this century, or in the first 18 months from January 2019---some designs behave differently from others.
There is to an extent a trade-off between these considerations: better positive skewness and better performance in certain trending scenarios may be obtained at the expense of worse average historical performance and vice versa. This necessitates subjective decision, and despite the great effort of systematic trading firms to claim that there is no discretion in the implementation of their systems---nowadays assisted by the smoke-screens of statistical theory and `machine learning'---inevitably there must be.

An incidental conclusion from reading \cite{Dao16} is that the SG CTA index is very easily replicated, giving the lie to the contentions, blithely trotted out by the CTA industry, that barriers to entry are so high, that the subject can only be understood by those with years of experience, that a cohort of PhDs are required to build strategies, and that proprietary execution algorithms are important---the last of these is clearly nonsense given the low speed at which the replicating strategy in \cite{Dao16} trades (see Figure~7 in that paper). 
In fact, rather than clever trade execution being important for momentum strategies, it is the reverse that is true: momentum is an important ingredient in trade execution, as over short time scales many financial time series exhibit momentum.  
Further, the relatively poor performance of the SG CTA index since the end of the Global Financial Crisis, together with the underlying simplicity of momentum trading, should make investors question whether CTAs' management fees are justified, as well as how big an asset allocation they should receive by comparison with standard investments in equities and fixed income.

A consequence of positive skewness is that the proportion of winning trades may well be negative \cite{Potters05}.
Small trading losses are common, but occasional big gains are produced when the strategy levers itself into a trend.  The longevity of trend-following funds suggests that this characteristic has served them well over the years, pointing to the conclusion that the oft-asked question ``What is your fraction of winning trades?'' is misleading.
The link between moments and proportion of winning trades can be formalised with the Cornish--Fisher expansion, which estimates for a random variable $Y$,
\[
\pr(Y>\ex[Y]) \approx \half - \frac{\kappa_3}{6\sqrt{2\pi}}
\]
where $\kappa_3$ is the coefficient of skewness of $Y$ (see note\footnote{Take for example the exponential distribution: the exact probability of exceeding the mean is $e\inv\approx 0.368$; the Gram-Charlier approximation gives $\half-2/(6\sqrt{2\pi})\approx0.367$, which is very close. On the other hand there may be considerable divergence for other distributions: for example the third moment may not even exist, or one may have a right-skewed distribution with negative third moment but positive higher odd moments. Nonetheless, the above approximation works `better than one might suppose'.}).

We show that the skewness of the returns depends strongly on their period\footnote{By the $\pd$-period trading return we mean the gain in P\&L between date $n$ and date $n+\pd$. This is to be distinguished very carefully from the \emph{market} return which simply means the change in price of the traded asset over that period. We are assuming a discrete time model with the time increment being 1 day.}, so that even if the one-day returns have no skewness, longer-period returns may be skewed.
This may at first seem curious, and arises because successive daily \emph{trading} returns are not independent. Thus it is possible to obtain a skewed distribution by adding non-skewed random variables, if those variables are appropriately dependent.

This paper is arranged as follows. We begin with linear strategies (\S\ref{sec:linear}) and give a complete exposition of `skew theory' for them. We show how to calculate this skewness as a function of the return period, by simple application of residue calculus.  The skewness of the $\pd$-period trading return depends on $\pd$: it rises to a maximum at a period proportional to the typical response-time of the trending indicator, and then drops as $\pd^{-1/2}$ (eqs.~ \ref{eq:skew_asymp}, \ref{eq:ema1skew}, \ref{eq:ema2mom2}/\ref{eq:ema2mom3}). We then test on some real data and find reasonable correspondence. Finally we analyse a particular hybrid linear strategy that is not pure momentum and derive a simple condition that ensures positive long-term skewness of returns.

An interlude on the option-like nature of trend-following ensues (\S\ref{sec:option}) which is a natural consequence of \S\ref{sec:linear}. 
This has since been treated in an excellent paper by Dao et~al.\ \cite{Dao16} who point out that in effect one is buying long-dated options and selling short-dated ones. This explains neatly why momentum strategies suffer badly from whipsawing, when short-date volatility is high, and perform well when the market moves steadily in one direction.
The effect is important because it causes momentum strategies to hold on to most of their previous profits during the periods where they are not making money. As pointed out by Till \& Eagleeye \cite{Till11}, this ``long-option behaviour'' distinguishes them from other strategies that tend to have higher Sharpe ratio, the implication being that the higher Sharpe is a form of remuneration for negative skewness. By the same token, as pointed out in a different context in \cite{Martin18a}, positive skewness can result in longer drawdown times than strategies with zero skew, depending on what period of history is being considered---so time spent in drawdown is not necessarily a good measure of strategy performance.

We then move on to nonlinear strategies in \S\ref{sec:nonlinear}.
An arbitrary nonlinear function of several momentum factors (of different speeds) would be very difficult to analyse, so we opt for nonlinearly transforming each momentum factor first, and then the position is a weighted sum of the transformed factors.
We extend the work on skewness of trading returns, studying the effect of the nonlinear transformation.
This analysis is primarily a matter of algebra, and we derive new results (\ref{eq:mom3};\ref{eq:Hkold},\ref{eq:Hk}) for the term structure of the skewness of trading returns. For some useful instances of the model (\ref{eq:Hk_ss},\ref{eq:Hk_rs},\ref{eq:Hk_ds}) we can evaluate these expressions in closed form, making for easy computation.
It turns out that the nature of the transformation is very important and can cause the positive skewness to disappear or even become negative.
For example, one simple transformation is the binary construction with a position of $+1$ or $-1$ according as the momentum factor is positive or negative---but, as we show here, there are good reasons to suppose that it is not optimal.  We conclude with some general remarks about the optimal design.

\begin{figure}[h!]
\begin{center}\begin{tabular}{c}
\scalebox{0.8}{\includegraphics*{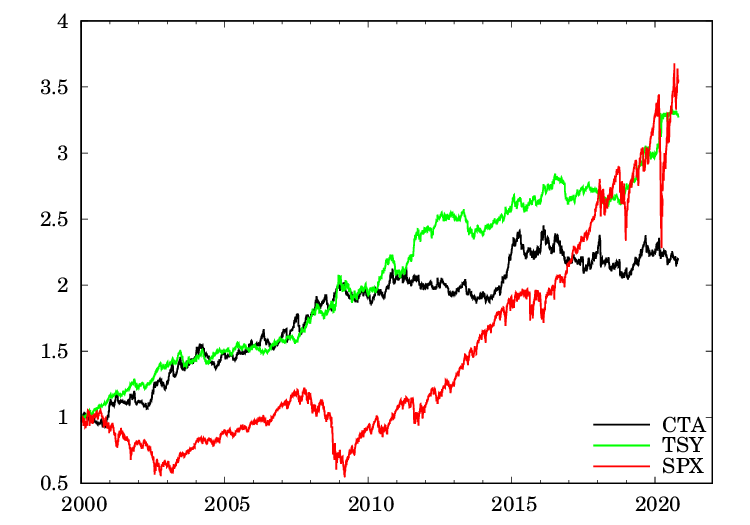}} \\
\scalebox{0.8}{\includegraphics*{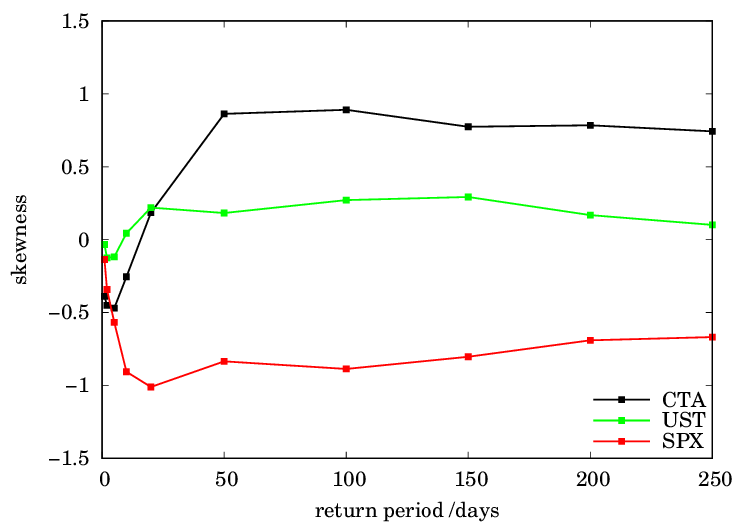}} \\
\end{tabular}\end{center}
\caption{\small Performance, or `total return', and skewness of market returns, for three indices: SG CTA index;  US Treasuries (7--10y bucket); equities (SPX). Data source for top plot: Bloomberg.}
\label{fig:0}
\end{figure}

\section{Basics and model setup}
\label{sec:basics}

We have already mentioned \emph{returns} and now formalise this notion. Simply, the return is either the change in price, or the relative change in price, the former being $X_{n+1}-X_n$ and the latter $(X_{n+1}-X_n)/X_n$ which is approximately the difference in $\ln X$ between the two time points. The latter can only be applied when prices are positive and is therefore inapplicable to asset classes such as interest-rate swaps\footnote{As a swap at inception has zero PV. If one wants to consider the momentum of the underlying swap rate, rather than the contract PV, then this is quite sensible, though not quite the same as the contract PV includes the effects of carry and rolldown, not just the changes in par swap rate. Nonetheless one should still use absolute changes for the obvious reason that interest rates can go negative; also the absolute variation in EUR and USD rates as examples has not been strongly spot-dependent over the last 25 years or so.}, but it is the most natural definition for equities, bond futures and most commodities. We say `most' commodities because a major upset occurred in the front WTI oil contract in April 2020 when it went negative \cite{Martin20b}. On the other hand the former definition is most natural for interest-rate futures. We use the former definition in the ensuing algebra but this choice is not critical to the theoretical development.

We define $U_{n+1}$ to be the return per unit volatility\footnote{Also known as the risk-adjusted return.} for the asset $X$ between time $n$ and $n+1$, i.e.
\begin{equation}
U_{n+1}= \frac{X_{n+1}-X_n}{\hat{\sigma}_n }, \qquad \hat{\sigma}_n = \ex_n[(X_{n+1}-X_n)^2]^{1/2}
\end{equation}
with $\ex_n$ denoting an expectation conditional on $\mathcal{F}_n$, the information known up to and including time $n$. The reason for dividing by $\hat{\sigma}$ is that we wish $U_n$ to be nondimensional and appropriately normalised.

Following the general principle that one should bet a number of contracts (or contract notional) inversely proportional to the contract volatility\footnote{Essentially to keep a reasonably constant level of risk on: if the same position is held while the volatility rises substantially, one is likely to break one's market risk limits. More formally this follows from stochastic control theory, see e.g.\ \cite[\S14]{Bjork98}. Writing $\mu$ for the drift and $\sigma$ for the volatility of the traded asset, the optimal position always emerges as $\mu/\sigma^2$ multiplied by a few factors that pertain to the exact setup (utility, etc). If the dimensionless trading signal $S$ is representative of the risk-adjusted return $\mu/\sigma$, the position is $S/\sigma$.}, we define the position in the asset to be $\sig_n/\hat{\sigma}_n$ at time $n$, where $\sig_n$ is to be a function of any or all of the $U$'s up to and including $U_n$.  One must have $\sig_n\in\mathcal{F}_n$, otherwise the strategy can cheat by looking ahead.
Clearly the P\&L arising from the period between time $n$ and time $n+1$ is $\sig_n U_{n+1}$. 

For much of this paper we assume that the risk-adjusted returns $U_n$ are i.i.d.\footnote{The raw returns need not be, because of stochastic volatility, which is another reason for dividing off by an estimate of the stdev of the asset return.} and of zero mean. Thus, we are studying the behaviour of the strategy under the assumption that it is not generating any expected return (Potters \& Bouchaud do the same).
We also assume that $U_n$ has zero third moment, so that its first three moments are 0,1,0. Thus although $U_n$ has no skewness, we are about to show that the trading return may have skewness. We make no further distributional assumptions about the $(U_n)$.

\notthis{
e make the following assumptions about the returns:
\begin{itemize}
\item[(i)] they are symmetrical in the sense that $\mom{U_n}{}=\mom{U_n}{3}=0$; 
\item[(ii)] $\mom{\sig U_{n+k}}=0$ for all $\sig\in\mathcal{F}_n$ and $k\ge1$ (this can be informally called independence, but is actually a weaker condition\footnote{Because we do not need to rule out the possibility that the magnitude of $U_n$ is dependent on previous values.}).
\item[(iii)] $\mom{\sig U_{n+k}^2}=\mom{\sig}\mom{ U_{n+k}^2}$ for all $\sig\in\mathcal{F}_n$ and $k\ge1$, another independence condition of sorts.
\end{itemize} 
(Actually (ii) implies the first condition of (i).)
} 

The period-$\pd$ trading return is defined as 
\begin{equation}
Y^{(\pd)}_n = \sum_{k=0}^{\pd-1} \sig_{n+k} U_{n+k+1}.
\end{equation}

The first moment of the trading return is clearly zero. The second moment is given by\footnote{$\mom{\cdot}$ denotes realisation average.}
\[
\bigmom{ (\sig_{0}U_{1} + \cdots + \sig_{\pd-1}U_\pd)^2 }.
\]
Let us consider this expectation on expansion as a product. The cross-terms all vanish because each contains a $U_{n+1}$ term multiplied by a term in $\mathcal{F}_n$. This leaves the squared terms, which give simply
\[
\pd \mom{\sig^2}
\]
(as $\mom{U^2}=1$). The proportionality in $\pd$ is a consequence of the trading returns being uncorrelated (note that we have \emph{not} said `independent').

The third moment of the trading return is given by
\[
\bigmom{ (\sig_{0}U_{1} + \cdots + \sig_{\pd-1}U_\pd)^3 }.
\]
Expanding this as a product, we obtain four types of terms:
\begin{itemize}
\item[(i)] $\sig_{n_1}U_{n_1+1}\sig_{n_2}U_{n_2+1}\sig_{n_3}U_{n_3+1}$ with $n_1<n_2<n_3$;
\item[(ii)] $\sig_{n}^2U_{n+1}^2\sig_{m}U_{m+1}$ with $n<m$; 
\item[(iii)] $\sig_{n}^2U_{n+1}^2\sig_{m}U_{m+1}$ with $n>m$;
\item[(iv)] $\sig_{n}^3U_{n+1}^3$.
\end{itemize}
The independence of the $U$'s and the assumptions about their moments show that (i), (ii) and (iv) all vanish. We are therefore left with (iii), which can be written
\begin{equation}
3 \sum_{0\le m<n <\pd} \mom{\sig_n^2\sig_m U_{m+1}}.
\label{eq:mom3gen}
\end{equation}
As $\sig_n^2$ may depend on $U_{m+1}$, this expression is not necessarily zero, though it \emph{must} be zero when $\pd=1$ (as the sum in (\ref{eq:mom3gen}) is empty: the cause is symmetry of the market returns).  The factor of 3 comes from the three ways of permuting the indices in (iii).

It is worth mentioning that if the third moment of the market returns is nonzero then this is likely to influence the skewness of the trading returns. `Risk markets' such as equities and credit typically have returns that have positive expected return but negative skewness. Positivity of the expected return will make a momentum signal have on average a positive allocation, that is, $\ex[\sig_n]>0$. The terms listed as type (iv) above will then cause the skewness of the short-term trading returns to be negative. In fact this is visible in Figure~\ref{fig:0}.

\section{Linear strategies}
\label{sec:linear}

\subsection{Notation and definitions}

We now specialise these results to linear strategies, by which we mean
\begin{equation}
\sig_n = \sum_{j=0}^\infty a_j U_{n-j}.
\label{eq:lin}
\end{equation}
Linear strategies have several advantages. They are easily constructed, for example through EMAs ($a_j\propto\alpha^j$) which can be implemented recursively: see Appendix~\ref{sec:MPR}.
They are also easily added, so that one can combine momentum of different periods (or even have negative weights on momentum of certain periods, so that one may capture counter-trending behaviour). Finally, analysis is reasonably straightforward, and the moments of the trading returns can be captured using the coefficients $(a_j)$ alone.

We shall need the autocovariance function of the impulse response:
\begin{equation}
R^a_k = \sum_{j=0}^\infty a_ja_{j+k}, \qquad k\ge0
\label{eq:Rak}
\end{equation}
and also the system function, i.e.\ the $z$-transform of the weights:
\begin{equation}
A(z)=\sum_{j=0}^\infty a_j z^{-j}, \qquad z\in\C.
\label{eq:A(z)}
\end{equation}
This is bounded for $|z|\ge1$.
A linear combination of EMAs always has a rational system function and its poles are usually a key part of the design and analysis. For a general account, refer to \cite{Haykin89}.

The simplest example is the single-EMA case, which we will call `EMA1': $a_j=\alpha^{j+1}$, and $A(z)=\frac{\alpha}{1-\alpha z\inv}$. 
This arises as the difference between the spot price and an EMA of past prices, and is used by Potters \& Bouchaud in \cite{Potters05}.
The decay-factor $\alpha$ is linked to the effective period of the EMA, $N$, by $\alpha=1-N\inv$, so that the EMA becomes progressively slower, or more highly smoothed, as $\alpha\to 1$.

Another important example is the difference of two expressions of this form, which we call `EMA2':
$a_j=\frac{\alpha^{j+1}-\beta^{j+1}}{\alpha-\beta}$ and $A(z)=\frac{1}{(1-\alpha z\inv)(1-\beta z\inv)}$.
This arises as the difference between two EMAs of prices, a common device in technical analysis\footnote{See for example \cite[\S9]{Herbst92} and also in many online articles on tea-leaf reading, e.g.\ {\tt www.stockcharts.com/school}.}. It has less day-to-day variation than EMA1, on account of being the difference of two smoothed prices, or equivalently a double (rather than single) EMA of the returns.

It is convenient to define a class of models that we call SPRZ (`simple poles, regular at zero'). The precise conditions are: $A(z)$ bounded in $|z|>1-\varepsilon$ for some $\varepsilon>0$; the only singularities to be simple poles; and $A(z)$ regular at the origin. These should be thought of as mild analytical conditions that enable the ready application of residue calculus; models with multiple poles can be understood as limiting cases of models with simple poles as the poles coalesce.

\subsection{Note on continuous time formulation}

\emph{This subsection may be omitted at a first reading.}

\newcommand{\dalpha}{\dot{\alpha}}
\newcommand{\dbeta}{\dot{\beta}}

The definitions of the EMA relate neatly to a continuous-time setting. For a process $X_t$ we can define any time-invariant linear system as
\[
\mathcal{K}[X]_t = \int_{-\infty}^t K(t-s) \, dX_s
\]
where $K$ is commonly known as the kernel. If $X_t$ is a unit Brownian motion then the variance of $\mathcal{K}[X]_t$ is
\[
\| \mathcal{K} \|^2 = \int_0^\infty K(t)^2 \, dt,
\]
which we call the square-norm.

An EMA1 is then the difference between $X$ and its exponentially-weighted moving average, which is an exponential smooth of the returns:
\begin{equation}
X_t - \int_{-\infty}^t \dalpha e^{\dalpha (s-t)} X_{s} \,ds =\int_{-\infty}^t e^{\dalpha (s-t)} dX_s
\label{eq:ctmom1}
\end{equation}
and its square-norm is $1/2\dalpha$ (of course $\dalpha>0)$.

An EMA2 is the difference of two of these, with $\dbeta>\dalpha$:
\begin{equation}
\int_{-\infty}^t \big(e^{\dalpha (s-t)}- e^{\dbeta(s-t)}\big) \,dX_{s}
\label{eq:ctmom2}
\end{equation}
and its square-norm is  $(\dalpha-\dbeta)^2 \big/ 2\dalpha\dbeta(\dalpha+\dbeta)$.
The limit $\dbeta\to\dalpha$ obviously only makes sense if we divide by $\dbeta-\dalpha$ first, giving
\begin{equation}
\int_{-\infty}^t (t-s) e^{\dalpha (s-t)} \,dX_{s}
\label{eq:ctmom2=}
\end{equation}
and its square-norm is $1/4\dalpha^3$. This can also be written as the EMA of $X$ minus the double-EMA (EMA of the EMA) of $X$. We call this `EMA2='.

An important practical aspect of continuous-time signals is the notion of path-length, defined for a process $Y_t$ to be 
\[
\frac{1}{T} \int_0^T |dY_t|.
\]
This is a concern because it relates to the rate at which money is lost in proportional transaction costs. Infinite path length results in an infinite rate of loss, unless the problem is obviated. For a Brownian motion the path length is infinite as $|dY_t|=O(\sqrt{dt})$. Now if $X_t$ is a Brownian motion and we pass it through a linear system of kernel $\mathcal{K}$, then what is the path length of $\mathcal{K}[X]_t$? 
We see that
\[
\int_{-\infty}^{t+dt} K(t+dt-s) \, dX_s - \int_{-\infty}^{t} K(t-s) \, dX_s  = K(0) \, dX_t +  dt \cdot \int_{-\infty}^t K'(t-s) \, dX_s 
\]
and the first term will generate infinite path length unless $K(0)=0$. The second term (without the $dt$) is Normally distributed of zero mean and variance $\|\mathcal{K}'\|^2$, so its expected absolute value is $(2\|\mathcal{K}'\|^2/\pi)^{1/2}$. The conclusion is that {\bf the path length is finite for EMA2, infinite for EMA1}. 
This does not rule out the use of EMA1, as the theory of trading under proportional transaction costs is reasonably well-established (see e.g.\ \cite{Martin11a,Martin12a,Martin14b} and references therein), but it does suggest that in trading systems EMA2 is preferable.

Now it may be advantageous to minimise the path-length, subject to two conditions: (i) the variance of the output is to be unity (as otherwise the solution would be $K\equiv0$), and (ii) the average lookback period is to be fixed (as otherwise we could take a normalised EMA2 and allow both speeds to tend to zero). The latter constraint can be implemented in various ways, but a felicitous one turns out to be
\begin{equation}
\mbox{ Minimise } \int_0^\infty K'(t)^2 \, dt 
\mbox { s.t. }
\int_0^\infty K(t)^2 \, dt = 1
\mbox{ and }
\int_0^\infty t^{-1} K(t)^2 \, dt = 1/\tau
\end{equation}
where $\tau$, of units time, is a given parameter.
One boundary condition is $K(0)=0$, and we require sensible behaviour at $\infty$.
This is a standard type of variational calculus problem and gives rise to the ODE
\begin{equation}
K''(t) + (\lambda + \mu t^{-1}) K(t) = 0
\end{equation}
where $\lambda,\mu$ are Lagrange multipliers.
It is an easy exercise to see that $te^{-\dalpha t}$ is one solution of this\footnote{Depending on the Lagrange multipliers. There an infinity of solutions, and the `sensible' ones are of the form polynomial $\times$ exponential. The polynomials in question are related to the Laguerre polynomials.}, and so in this particular sense EMA2= is an optimal choice of momentum filter.

\subsection{Second and third moments}

It is immediate that the second moment of the $\pd$-period trading return is $\pd R^a_0$. For the third moment, we have to find the $U_{m+1}U_j$ term ($j\ne m+1$) in $\sig_n^2$ in the expression (\ref{eq:mom3gen}). This is
\[
2 \sum_{k=0,k\ne n-m}^\infty a_{n-m}a_k U_{m+1}U_{n-k+1}.
\]
This now has to be multiplied by $3\sig_mU_{m+1}$ and the expectation taken. Thus it is necessary to look for any overlap between $U_{n-k+1}$ and $\sig_m$, and so in the $k$-summation we only need terms with $k\ge n-m$, and exclude the others. The resulting expression emerges as
\begin{equation}
 3 \sum_{0\le m < n <\pd} 2 \sum_{k=n-m}^\infty a_{n-m-1} a_k a_{m-n+k} 
=
6 \!\!\! \sum_{0\le m <n <\pd} \!\!\! a_{n-m-1} R^a_{n-m} 
=
6 \sum_{k=1}^{\pd-1} (\pd-k) a_{k-1} R^a_k.
\label{eq:mom3lin}
\end{equation}

\notthis{
The first, second and third moments of the trading return are, respectively,
\begin{equation}
\mom{Y}= 0 ; \qquad \mom{Y^2} = \pd A_0 ; \qquad \mom{Y^3} =  6\sum_{k=1}^{\pd-1} (\pd-k) a_{k-1} A_k
\end{equation}
} 

If we understand a pure momentum, or trend-following, strategy to be one in which all the $(a_j)$ are positive, then by (\ref{eq:mom3lin}) the trading returns must be positively skewed. By the same token a counter-trending strategy, with the $a$'s negative, has a negatively skewed return distribution even if the market returns are symmetrical. 

We can $z$-transform (\ref{eq:mom3lin}), i.e.\ multiply by $z^{-\pd}$ and sum from $\pd=1$ to $\infty$, to get
\begin{equation}
G_3(z) = \frac{6z}{(z-1)^2} \sum_{k=1}^\infty a_{k-1} R^a_k z^{-k}.
\end{equation}
From the presence of a double pole in $G_3(z)$ at $z=1$ we deduce that the third moment is asymptotically $6\pd\sum_{k=1}^\infty a_{k-1} R^a_k$ as $\pd\to\infty$.
Recalling that the second moment is linear in $\pd$, we deduce that as a function of the return period $\pd$ the skewness starts from zero, reaches an extremum somewhere and decays as $\pd^{-1/2}$.

The $\pd^{-1/2}$ asymptotic is, intriguingly, the same as that observed in L\'evy processes. However, the origin of the skewness is completely different. With a L\'evy process, it arises because the one-period returns are asymmetrical but independent. Here, they are symmetrical but not independent!

\subsection{Further analysis of the third moment}

The asymptotic third moment (without the $6\pd$ prefactor) is $\sum_{k=1}^\infty a_{k-1} R^a_k$ which is also equal to 
\[
\frac{1}{2\pi\I} \oint_{|z|=1} A(z)A(z\inv)^2 \, dz
\]
(to see this, write the integrand as a product of Taylor series; the integral pulls out the $z\inv$ term). This expression can be calculated using residue calculus, if we restrict ourselves to the SPRZ case, as
\[
\sum_j \rho_j A(\alpha_j\inv)^2.
\]
Meanwhile the second moment is $\pd R^a_0$, and
\begin{equation}
R^a_0 = \frac{1}{2\pi\I} \oint_{|z|=1}  A(z)A(z\inv) z\inv\,dz
=A(0)^2 + \sum_j \rho_j \alpha_j\inv A(\alpha_j\inv).
\label{eq:mom2lin}
\end{equation}
Collecting the results together, we deduce that the skewness of $\pd$-period trading returns, for large $\pd$, is
\begin{equation}
\kappa_3^{(\pd)} \sim
\frac{
6 \sum_j \rho_j A(\alpha_j\inv)^2
}{
\left(A(0)^2 + \sum_j \rho_j \alpha_j\inv A(\alpha_j\inv)\right)^{3/2} \pd^{1/2}
}.
\label{eq:skew_asymp}
\end{equation}

In the EMA1 case we immediately obtain
\[
\kappa_3^{(\pd)} \sim \frac{6\alpha}{(1-\alpha^2)^{1/2} \pd^{1/2}} \sim 3\sqrt{2} \left( \frac{N}{\pd} \right)^{1/2}
\]
where the right-hand expression is obtained by assuming that $N=(1-\alpha)\inv$ is not small.
In the EMA2 case, the poles are at $\alpha$, $\beta$ and are of residue $\alpha^2$, $-\beta^2$, and the result is, after a little algebra,
\[
\kappa_3^{(\pd)} \sim
 \frac{6(\alpha+\beta)(1-\alpha\beta)^{1/2}}{(1-\alpha^2)^{1/2}(1-\beta^2)^{1/2}(1+\alpha\beta)^{1/2}\pd^{1/2}}
\sim 3\sqrt{2} \left(\frac{N_\alpha+N_\beta}{\pd}\right)^{1/2}
.
\]

We can also return to (\ref{eq:mom3lin}) to get the exact third moment, not just the long-term asymptotic. To do this, we write (\ref{eq:mom3lin}) in terms of $A(z)$, as
\[
\frac{6}{(2\pi\I)^3} \sum_{k=1}^{\pd-1} (\pd-k) \sum_{j=0}^\infty \oint \oint \oint A(y)y^{j-1} A(z) z^{j+k-1} A(w)w^{k-2} \, dw\, dy\,dz
\]
in which the contours for $w$- and $y$-integrals are of radius $1-\varepsilon$ and the contour for $z$ is just $|z|=1$ (the need for this will presently become apparent). The $j$-summation and the $y$-integral can be done immediately (the placement of the contours causes $|yz|<1$, which is necessary for convergence of the sum; in doing the $y$-integral, expand the contour out to $\infty$ and pick up the residue at $y=1/z$ on the way). Next do the $k$-summation using the identity
\[
\sum_{k=1}^{m-1} (m-k)r^{k-1} \equiv \frac{r^m-1+m(1-r)}{(1-r)^2}
\]
to arrive at
\[
\frac{6}{(2\pi\I)^2} \oint \oint A(z\inv)  A(z) A(w) \frac{(wz)^\pd-1 +\overbrace{\pd(1-wz)}^\ast}{(1-wz)^2}  w^{-1} \, dw\, dz.
\]
The marked term exactly generates the large-$\pd$ result we have already obtained, once the $w$-integral is done (again, by expanding the contour out to $\infty$ and picking up the residue at $w=1/z$ on the way).
The remaining part can be calculated by collapsing the $w$-contour around all the singularities inside the unit circle (note that no singularity arises from the $(1-wz)^2$ term in the denominator, as $|wz|<1$), and then collapsing the $z$-contour. In the SPRZ case, we finally obtain the third moment as
\begin{equation}
6\pd \sum_j \rho_j A(\alpha_j\inv)^2 - 6A(0) \sum_j \rho_j A(\alpha_j\inv) - 6\sum_{j,k} \rho_j\rho_k \alpha_j\inv A(\alpha_k\inv) \frac{1-\alpha_j^\pd\alpha_k^\pd}{(1-\alpha_j \alpha_k)^2}.
\label{eq:mom3pol}
\end{equation}

\begin{figure}[h!]
\begin{center}\begin{tabular}{c}
(a) \scalebox{0.8}{\includegraphics*{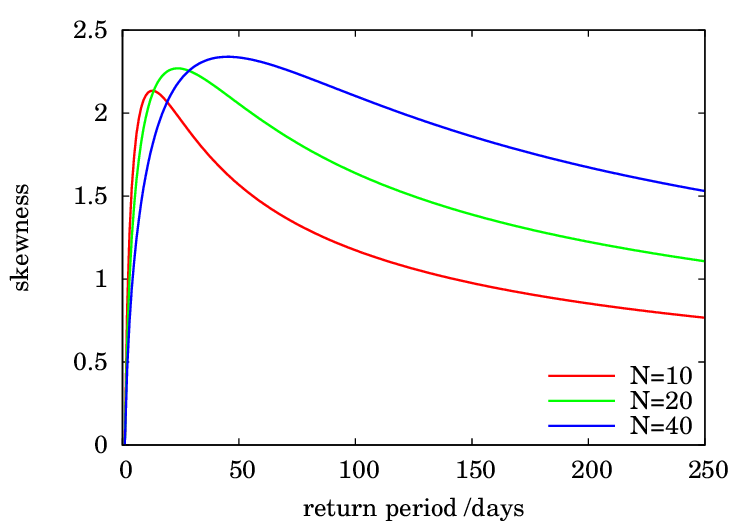}} \\
(b) \scalebox{0.8}{\includegraphics*{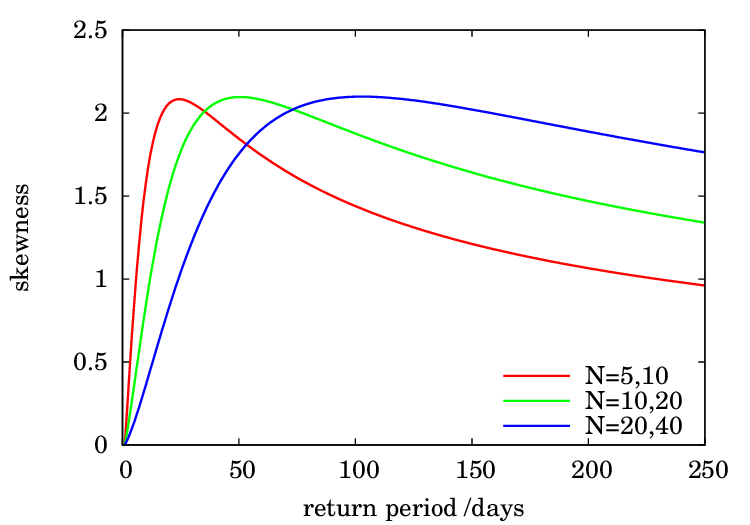}} \\
\end{tabular}\end{center}
\caption{\small Skewness of trading returns, as a function of period, for (a) EMA1 type model, (b) EMA2 type model. Note the characteristic shape.}
\label{fig:1}
\end{figure}

For EMA1, the exact expression for the skewness is therefore
\begin{equation}
\kappa_3^{(\pd)}=
\frac{6\alpha}{(1-\alpha^2)^{1/2}\pd^{1/2}} \left( 1- 
\frac{1-\alpha^{2\pd}}{1-\alpha^2} \pd^{-1} \right).
\label{eq:ema1skew}
\end{equation}
This rises from zero to a peak and then rolls off as $O(\pd^{-1/2})$ (see Figure \ref{fig:1}a). The maximum skew\footnote{Unless $N$ is small we can approximate (\ref{eq:ema1skew}) as $(3/\sqrt{2x^3})(e^{-2x}-1+2x)$, with $x=\pd/N$; the maximum of this function is $\approx2.41$ and occurs at $x\approx1.07$.} is roughly 2.1--2.4, and occurs for period $\pd\approx1.1N$ (recall $\alpha=1-N\inv$).

For EMA2, the exact expressions for the second and third moments are
\begin{eqnarray}
\mu_2^{(\pd)} = && \frac{\pd (1+\alpha\beta)}{(1-\alpha\beta)(1-\alpha^2)(1-\beta^2)} \label{eq:ema2mom2}
\\
\mu_3^{(\pd)} = && \frac{6\pd (\alpha+\beta)(1+\alpha\beta)}{(1-\alpha\beta)(1-\alpha^2)^2(1-\beta^2)^2} \label{eq:ema2mom3} \\
&& \; +\, \frac{6\alpha^3(1-\alpha^{2\pd})}{(\alpha-\beta)^2(1-\alpha^2)^3(1-\alpha\beta)} + \frac{6\beta^3(1-\beta^{2\pd})}{(\alpha-\beta)^2(1-\beta^2)^3(1-\alpha\beta)} \nonumber \\
&& \; -\, \frac{6\alpha\beta^2(1-\alpha^\pd\beta^\pd)}{(\alpha-\beta)^2(1-\beta^2)(1-\alpha\beta)^3} - \frac{6\alpha^2\beta(1-\alpha^\pd\beta^\pd)}{(\alpha-\beta)^2(1-\alpha^2)(1-\alpha\beta)^3} \nonumber
\end{eqnarray}
and $\kappa_3^{(\pd)}=\mu_3^{(\pd)}\big/\big(\mu_2^{(\pd)}\big)^{3/2}$ as usual.
This is qualitatively similar to EMA1 (see Figure \ref{fig:1}b).  The maximum skew is around 2.1, and occurs at period $\pd\approx1.7(N_\alpha+N_\beta)$, provided $N_\alpha$ and $N_\beta$ are not too far apart. In the extreme case where either of the $N$'s is equal to 1, we recover EMA1.
The limit $\beta\to\alpha$ is well-behaved, but algebraically messy and omitted here.

In essence, (\ref{eq:mom3pol}) telescopes the various geometric series that are implicit in the calculation of (\ref{eq:mom3lin}), and allows it to be done with an amount of computational effort independent of $\pd$.

\subsection{Empirical results}

\begin{figure}[h!]
\begin{center}\begin{tabular}{c}
(a) \scalebox{0.8}{\includegraphics*{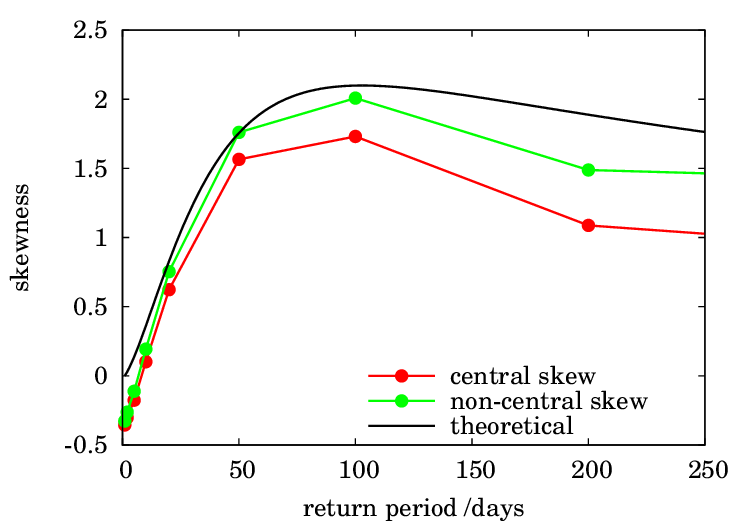}} \\
(b) \scalebox{0.8}{\includegraphics*{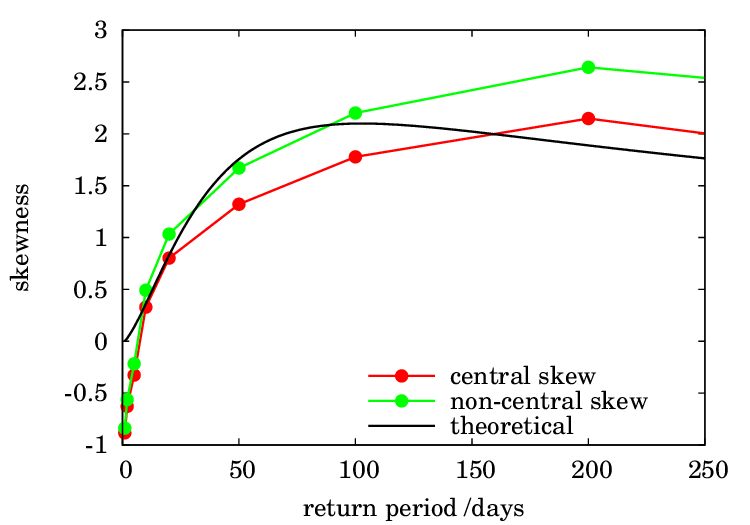}} \\
\end{tabular}\end{center}
\caption{\small Skewness of trading returns, as a function of period, for (a) CHFUSD, (b) S\&P500 futures; theoretical result also shown. $N=20,40$.}
\label{fig:2}
\end{figure}

For a demonstration using real data we use two datasets: the CHFUSD futures and the S\&P500 futures\footnote{Bloomberg: {\tt SF1 Curncy} and {\tt SP1 Index}. These are the front contracts, rolled 10 days before expiry to create generic series. Data range: 01-Jan-90 to 31-Dec-09.}.
For risk-adjusting the returns we use a 20-day EMA of squared price changes to estimate the volatility ($\hat{\sigma}_n$ in the definition of $U_n$).
We are using an EMA2 with $N=20,40$.

It is worth recalling the assumptions that we made in deriving our formulae: (i) independence of the risk-adjusted returns $(U_n)$; (ii) symmetry of their distribution up to the third moment. In practice the first clearly does not hold, because it implies that momentum strategies do not generate positive expected return, whereas the evidence is that on average they do. That means that when we examine real data, the observed skewness of returns may well not equal the theoretical result, by virtue of the mean being different. We therefore plot the central skew (third central\footnote{Central moment = moment about the mean.} moment divided by $\frac{3}{2}$ power of the second central moment---the usual definition) and also the `non-central' skew (third moment about zero divided by $\frac{3}{2}$ power of the second moment about zero). If the effect of trending is to generate a slightly positive expected return but keep the other moments roughly equal, then the non-central skew will be fractionally higher than the central skew. As to (ii), we know that equity markets occasionally have very negative returns.

Figure~\ref{fig:2} shows the results for the two markets, superimposing also the theoretical result from Figure~\ref{fig:1}b. In spite of the deficiencies in the modelling assumptions the agreement is not bad and the general shape is right. The short-term skewness for the equity market is nonzero because of the asymmetry of the market returns; the higher long-term skewness is best ascribed to the particularly good trending behaviour in the mid-1990s generating high trading returns.  
The skewness of the trading returns is far higher than that of the underlying markets (i.e.\ of the $U_n$'s): the latter is (to within 0.1) typically about 0.0 for CHFUSD and $-0.2$ for S\&P500. This shows that the skewness comes entirely from the momentum strategy. 
The Gram-Charlier formula for the probability of exceeding zero is modified to $\Phi(\mathfrak r)-\kappa_3/(6\sqrt{2\pi})$ in the presence of nonzero first cumulant (expected return); here $\mathfrak r=\kappa_1\big/\kappa_2^{1/2}$ is the Sharpe ratio and $\Phi$ is the Normal c.d.f. For horizons $\pd$ in the range 100--200 days the Sharpe ratio of each is roughly
  +0.2 and the skewness is around 2, so this gives the probability of exceeding zero as about 0.45, which corresponds well with the empirical value---note that it is less than one-half.

\subsection{Hybrid linear models}

\begin{figure}[h!]
\begin{center}\begin{tabular}{c}
\scalebox{0.8}{\includegraphics*{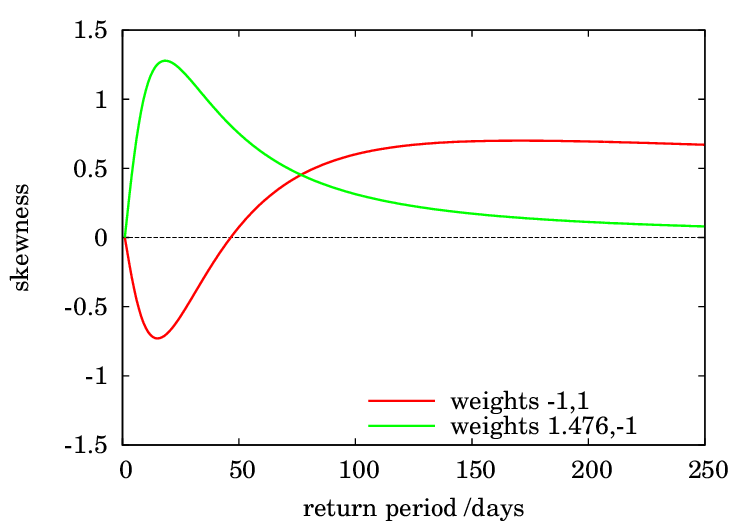}} \\
\end{tabular}\end{center}
\caption{\small Skewness of trading returns, as a function of period, for hybrid model with both trending and counter-trending behaviour, in two cases.}
\label{fig:3}
\end{figure}

Suppose that a strategy has a trend-following and a counter-trending characteristic, as would happen if its weights were obtained from a linear combination of EMA2's, with opposite signs. It may be desirable to ensure that the long-term skewness remains positive, as this is associated with longevity of the strategy. There are two situations in which this arises. In what is basically a trend-following strategy, it is desired:
\begin{itemize}
\item [(i)]
to make small bets on short-term reversion  without this upsetting the behaviour if a longer-term trend occurs;
\item [(ii)]
to make a small bet against very long-term trends on the supposition that what goes up must eventually come down (or vice versa), provided this bet is not too large.
\end{itemize}
In the first case the weights on most recent returns will be negative; in the second, it is the weights on the distant past that will be negative. The idea is to make sure that they are not too negative, in a sense to be made precise.

We have a model of the form
\[
A(z) = \frac{\lambda_F (\alpha_F-\beta_F)}{(1-\alpha_F z\inv)(1-\beta_F z\inv)} + \frac{\lambda_S (\alpha_S-\beta_S)}{(1-\alpha_S z\inv)(1-\beta_S z\inv)}  
\]
where $\lambda_F$ and $\lambda_S$ are the multipliers on the fast and slow components. Positive asymptotic skewness is ensured by (\ref{eq:skew_asymp}):
\begin{equation}
\sum_j \rho_j A(\alpha_j\inv)^2 > 0
\label{eq:poscdn}
\end{equation}
where there are now four poles $\alpha_1=\alpha_F$, $\alpha_2=\beta_F$, $\alpha_3=\alpha_S$, $\alpha_4=\beta_S$. Thus
\[
\rho_1 = \alpha_F^2\lambda_F, \qquad
A(\alpha_1\inv) = \frac{\lambda_F(\alpha_F-\beta_F)}{(1-\alpha_F^2)(1-\beta_F\alpha_F)} + \frac{\lambda_S(\alpha_S-\beta_S)}{(1-\alpha_S\alpha_F)(1-\beta_S\alpha_F)}
\]
and similarly for the other three.
The LHS of (\ref{eq:poscdn}) is a homogeneous cubic in $\lambda_F,\lambda_S$, which will factorise as
\[
\mathcal{P}(\lambda_F,\lambda_S) = (\lambda_F - \zeta_1 \lambda_S)(\lambda_F - \zeta_2 \lambda_S)(\lambda_F - \zeta_3 \lambda_S)
\]
where the $\zeta$'s are functions of the four poles. It is possible to identify the coefficients of $\lambda_F^3$, $\lambda_F^2\lambda_S$, $\lambda_F\lambda_S^2$, $\lambda_S^3$ as functions of the poles, then evaluate them and factorise the cubic by the Cardano-Tartaglia formula. However for practical purposes one might just as well write a numerical routine for LHS(\ref{eq:poscdn}) and find the roots $\zeta_i$ numerically. One root $\zeta_1$ has to be real, and the other two are likely to be complex because we expect $\mathcal P$ to be strictly increasing in $\lambda_F$ and in $\lambda_S$: raising either weight should enhance the trending behaviour and hence the asymptotic skewness. 

As a particular example, let $N_{\alpha_F}=5$, $N_{\beta_F}=10$, $N_{\alpha_S}=20$, $N_{\beta_S}=40$. Then $\zeta_1\approx-1.476$, and the condition for positive asymptotic skewness is simply\footnote{Because the other two terms in $\mathcal{P}$ multiply to give a quadratic that is always positive, and hence of no consequence.}
\[
\lambda_F + 1.476\, \lambda_S > 0.
\]
This being so, it is easily incorporated into an optimisation as a `style constraint'. Figure \ref{fig:3} shows the results for two examples, (i) $\lambda_F=-1$, $\lambda_S=1$, so the short-term behaviour is counter-trending and generates negative skewness; (ii) the critical case $\lambda_F=1.476$, $\lambda_S=-1$, where now just enough long-term counter-trending behaviour is added to make the asymptotic skew zero at leading order. These exemplify the cases (i), (ii) discussed above. 
The results were obtained using (\ref{eq:mom3pol}) again, which is not laborious despite there being four poles (so that the double summation has sixteen terms): it is preferable to Monte Carlo simulation, which even with a few hundred thousand simulations generates noticeable uncertainty.

\section{Option-like nature of trend-following}
\label{sec:option}

As pointed out in \cite{Till11}, trend-following strategies are often thought to have a long-option-type payoff on account of the positive skewness. For linear strategies this can be formalised as follows. The $\pd$-period trading return is
\[
Y^{(M)}_n = \mathbf{ u' \Gamma  u} , \qquad
\mathbf{u} = \begin{bmatrix} U_{n+\pd} & U_{n+\pd-1} & \cdots \end{bmatrix}'
\]
where the symmetric matrix $\mathbf{\Gamma}$ is given by
\begin{equation}
\mathbf{\Gamma} = \half \begin{bmatrix} 
0 & a_0 & a_1 & a_2 & \cdots \\ a_0 & \ddots & \ddots & \ddots & \ddots & \ddots \\ a_1 & \ddots & 0 & a_0 & a_1 & a_2 & \cdots \\
a_2 & \ddots & a_0 & 0 & 0 & 0 & \cdots \\ \vdots & \ddots & a_1 & 0 & 0 & 0 & \cdots \\ & \ddots & a_2 & 0 & 0 & 0 & \cdots \\ & & \vdots & \vdots & \vdots & \vdots & \ddots 
\end{bmatrix}
\demarcate.
\label{eq:Gamma}
\end{equation}
The moments of $Y^{(\pd)}_n$ relate to the spectrum of $\mathbf\Gamma$, and direct calculation reveals
\[
\bigmom{Y^{(\pd)}_n} = \tr (\mathbf\Gamma) = 0, \quad
\bigmom{\big(Y^{(\pd)}_n\big)^2} = 2\, \tr (\mathbf\Gamma^2) , \quad
\bigmom{\big(Y^{(\pd)}_n\big)^3} = 8\, \tr (\mathbf\Gamma^3) .
\]
Writing $\mathbf\Gamma$ in terms of its eigenvalues $\gamma_j$ and normalised eigenvectors $\mathbf e_j$, 
we have an expression that is a weighted sum (weights adding to zero) of `orthogonal quadratic bets', i.e.\ squared linear combinations of returns, which are like straddle payoffs but have constant convexity:
\begin{equation}
Y^{(\pd)}_n = \sum_j \gamma_j (\mathbf e_j \cdot \mathbf u)^2 .
\label{eq:sumsq}
\end{equation}
Now $\tr (\mathbf\Gamma ^r)=\sum_j \gamma_j^r$, so the moments of $Y^{(\pd)}_n$ relate to the moments of the eigenvalue distribution. It is easy to see the rank of $\mathbf\Gamma$ is $\le 2\pd$ (and is $ \pd+1$ in the EMA1 case as then the rows after the $\pd$th are linear multiples of each other), which limits the number of nonzero eigenvalues to $2\pd$.
The interpretation of all this is that a positively skewed strategy has a small number of large positive eigenvalues and a larger number of smaller negative ones. This generates a small number of large positive-convexity bets and a larger number of smaller negative-convexity bets, which is where the positive skewness comes from. Dao et~al.\ \cite{Dao16} make the same point, but emphasise the important point that the positive-convexity bets are long-dated options and the negative-convexity bets short-dated ones. Thus in situations where the long-term volatility is elevated and the short-term volatility is low, momentum strategies work well.

\subsection{*Full distribution of trading returns}

\emph{This subsection may be omitted at a first reading.}

If we want to know the full distribution of trading returns, we need to make an assumption about the full distribution of the market returns, whereas until now we have only used the first three moments. 

We can use the ideas of the previous section to compute the full distribution of trading returns, exactly as is done by Acar \cite[Ch.3]{Acar92} using generating functions.
The moment-generating function of the $\pd$-period return is
\[
F_\pd(s) := \bigmom{\exp (s Y^{(\pd)} )} = \bigmom{\exp \big(s(\sig_0U_1+\cdots+\sig_{\pd-1}U_\pd)\big)}.
\]
Let us assume that the $(U_n)$ are Normally distributed. Then for a linear model
\[
F_\pd(s) = [\det (\mathbf I - 2s \mathbf \Gamma)]^{-1/2}
\]
with $\mathbf \Gamma$ as above. (See e.g.\ \cite{Feuerverger00} for details on quadratic transformations of Normal variables.)

Before proceeding further we should note that the above expression is not very helpful because it requires the manipulation of $\mathbf{\Gamma}$ which is an infinite matrix. Let us therefore evaluate \emph{ab initio} the expression  
\[
\bigmom{\exp \big(s(\sig_0U_1+\cdots+\sig_{\pd-1}U_\pd)\big)}.
\]
Conditioning on $(U_1,\ldots,U_\pd)$, effectively fixing those values, we have (inside the exponential) a linear combination of $U_0,U_{-1},\ldots$, added to another expression that is a function of ${(U_j)}_{j=1}^\pd$ only. In the first part the coefficients are
\[
\begin{array}{ll}
U_0 :& s(a_0U_1+a_1U_2+\cdots+a_{\pd-1}U_\pd) \\
U_{-1} :& s(a_1U_1+a_2U_2+\cdots+a_{\pd-2}U_\pd)
\end{array}
\]
and so on. These variables can then be integrated out to give the expression
\[
\exp \Bigg\{ \shalf s^2 \sum_{l=-1}^\infty \left( \sum_{k=1}^\pd a_{k+l}U_k\right)^2 \Bigg\}
= 
\exp \Bigg\{ \shalf s^2 \sum_{l=-1}^\infty \sum_{j,k=1}^\pd a_{j+l}a_{k+l} U_j U_k \Bigg\}.
\]
The other part of the expression depends on the $U$'s through pairwise products, i.e.\ $U_1U_2$, etc, and is easily seen to be
\[
\exp \Bigg\{ s \sum_{1\le j<k \le \pd} a_{k-j-1} U_jU_k \Bigg\} .
\]
Next we multiply these two expressions to obtain
\[
F_\pd(s) = \bigmom{\exp \mathbf{u}'\mathbf{\widehat{G}}(s) \mathbf{u} },
\qquad
\mathbf{u} = \begin{bmatrix} U_{1} & U_{2} & \cdots & U_\pd \end{bmatrix}',
\]
with
\begin{equation}
\widehat{G}_{jk}(s) = \shalf s^2 \sum_{l=-1}^\infty a_{j+l}a_{k+l} + \shalf s \mathbf{1}_{j \ne k} a_{|k-j|-1}.
\end{equation}
Finally integrate $\mathbf u$ out, to get the more manageable
\begin{equation}
F_\pd(s)= [\det (\mathbf I - 2\mathbf{\widehat{G}}(s))]^{-1/2}
\end{equation}
instead---the dimension of $\mathbf{\widehat{G}}$ is just $\pd$. Again the infinite summation in the expression for $\widehat{G}_{jk}$ can be done easily enough by residue calculus in the SPRZ case.

The moments and cumulants can now be obtained by performing a Taylor series of, respectively, $F_\pd(s)$ and $\log F_\pd(s)$ around the origin, using standard identities for expanding the determinants. The derivation is laborious, however. To obtain an approximation to the distribution of $Y^{(\pd)}$, we may use Fourier transform inversion or, as discussed in 
\cite{Feuerverger00}, saddlepoint methods, which work well on this problem (see also \cite{Martin11b} for a general discussion).

\section{Nonlinear strategies}
\label{sec:nonlinear}

\notthis{
We define by $U_{n+1}$ the return per unit volatility\footnote{Also known as the risk-adjusted or vol-adjusted return.} for the asset $X$ between time $n$ and $n+1$, i.e.
\[
U_{n+1}= \frac{X_{n+1}-X_n}{\hat{\sigma}_n }, \qquad \hat{\sigma}_n = \ex_n[(X_{n+1}-X_n)^2]^{1/2}
\]
($\ex_n$ denoting an expectation conditional on $\mathcal{F}_n$). The reason for dividing by $\hat{\sigma}$ is that we wish $U_n$ to be nondimensional and appropriately normalised\footnote{There is a minor technicality though: $\hat{\sigma}^2_n$ is not as written the expected squared return, but only an `ex ante' \emph{estimate} of it, usually coming from some simple procedure such as an EMA of previous squared returns. Thus how well-normalised the $(U_n)$ are depends on how good this estimation is, though in practice it is not a significant issue.}.
} 

As in the first part of the paper we write the position as $\sig_n/\hat{\sigma}_n$. 
Clearly the P\&L arising from the period between time $n$ and time $n+1$ is $\sig_n U_{n+1}$.
We define a \emph{momentum factor} $V_n$ to be a moving average of the $U$'s, with weights $(a_j)_{j=0}^\infty$. The position, however, will no longer be simply proportional to $V_n$, but instead transformed using a nonlinear `activation\footnote{Following neural network parlance.} function' $\psi$:
\begin{equation}
V_n = \sum_{j=0}^\infty a_j U_{n-j}; \qquad \sig_n = \psi(V_n).
\label{eq:MA}
\end{equation}
We use the autocovariance function $R^a_k$ as before but now for convenience we stipulate that $V_n$ have unit variance, which is to say $R^a_0=1$. Then $|R^a_k|\le1$ (by Cauchy-Schwarz), and the inequality will be strict in all practical examples. Importantly, when designing the activation function $\psi$ we know that the typical scale of variation of its input is unity.

In the EMA1 model, we have
\[
a_j = (1-\alpha^2)^{1/2} \alpha^{j}, \qquad R^a_k = \alpha^k,
\]
differing trivially from what we had in \S\ref{sec:linear} by a scaling factor so that $R^a_0=1$.
For EMA2,
\[
a_j= \left(\frac{(1-\alpha^2)(1-\beta^2)(1-\alpha\beta)}{1+\alpha\beta}\right)^{1/2} \frac{\alpha^{j+1}-\beta^{j+1}}{\alpha-\beta},
\qquad
R^a_k = \frac{\alpha^{k+1}(1-\beta^2)-\beta^{k+1}(1-\alpha^2)}{(\alpha-\beta)(1+\alpha\beta)}.
\]
If $\beta=\alpha$ then this is well-defined and is simply a double-EMA of the returns.

This completes our description of the model setup: in summary, the position taken is a nonlinear transformation of smoothed vol-adjusted returns, divided by the volatility of the underlying. It is specified by the EMA parameter(s) and the function $\psi$. A general system can then be built by taking a linear combination of these models with different speeds.

\subsection{Skewness of nonlinear models}

We make the same assumptions as before about the nature of the risk-adjusted returns $U_n$, that is they are i.i.d.\ and their first three moments are 0,1,0. 
Then the second moment of the trading return is
\begin{equation}
\bigmom{ \big(Y^{(\pd)}_n\big)^2 } 
=\pd \mom{\sig^2}
\label{eq:mom2}
\end{equation}
and the third moment is
\begin{equation}
\bigmom{ \big(Y^{(\pd)}_n\big)^3 } 
=3 \sum_{k=1}^{\pd-1} (\pd-k) H_k, \qquad H_{n-m} = \mom{\sig_n^2\sig_m U_{m+1}}.
\label{eq:mom3}
\end{equation}

Whereas in \S\ref{sec:linear} we evaluated this summation directly, we can no longer do this and therefore take a slightly more roundabout route. Note first that the triple $(\sig_n^2,\sig_m,U_{m+1})$ is a simple function of the triple $(V_n,V_m,U_{m+1})=(Z_1,Z_2,Z_3)$ say, which has a trivariate Normal distribution with covariance matrix
\[
\mathbf{\Sigma} = \begin{bmatrix} 1 & R^a_{n-m} & a_{n-m-1} \\ R^a_{n-m} & 1 & 0 \\ a_{n-m-1} & 0 & 1\end{bmatrix}.
\]
The determinant and inverse of $\mathbf{\Sigma}$ are
\[
\Delta=1-a_{n-m-1}^2-R_{n-m}^2, \qquad
\mathbf{\Sigma}\inv = \frac{1}{\Delta} \begin{bmatrix} 1 & -R^a_{n-m} & -a_{n-m-1} \\ -R^a_{n-m} & 1-a_{n-m-1}^2 & a_{n-m-1}R^a_{n-m} \\ -a_{n-m-1} & a_{n-m-1}R^a_{n-m} & 1-(R^a_{n-m})^2 \end{bmatrix}.
\]
Writing the expectation $\mom{\sig_n^2\sig_m U_{m+1}}$ as an integral we get\footnote{Notation $\bz$ just means the column vector $(z_1,z_2,z_3)'$.}
\[
H_{n-m} = \frac{1}{(2\pi)^{3/2}\Delta^{1/2}} 
\iiint\limits_{\hspace{0mm}-\infty}^{\hspace{6mm}\infty} 
\exp(-\shalf \bz' \mathbf{\Sigma}\inv \bz) 
\psi(z_1)^2 \psi(z_2) z_3
 \,dz_1\, dz_2\, dz_3
\]
and then integrate over $z_3$ (effectively, integrating $U_{m+1}$ out). After a fair amount of algebra, the whole lot tidies up to give an expression as an expectation over the distribution of $(Z_1,Z_2)$ which is bivariate Normal with unit marginals and correlation $\rho=R^a_k$:
\begin{eqnarray}
H_k &=& 2a_{k-1} \left. \bigmom {\psi(Z_1)^2 \psi(Z_2) \frac{Z_1-\rho Z_2}{2(1-\rho^2)} } \right|_{\rho=R^a_k} \label{eq:Hkold} \\
 &=& 2a_{k-1} \left. \bigmom {\psi(Z_1)\psi'(Z_1) \psi(Z_2) } \right|_{\rho=R^a_k} ,
\label{eq:Hk}
\end{eqnarray}
where the last line follows from integrating by parts w.r.t.\ $Z_1$, assuming differentiability of $\psi$.

In the linear case, $\psi(z)=z$, we have 
$
H_k=2 a_{k-1}R^a_k 
$,
and hence the third moment is $6\sum_{k=1}^\pd (\pd-k)a_{k-1}R^a_k$ as we saw earlier.
When $\psi$ is not linear, we have to calculate $H_k$ for each $k$ and then do the sum by explicit calculation. 
That said, it is possible to find functions $\psi$ of the `right shape' for which the double integral implicit in (\ref{eq:Hk}) can be done in closed form. We discuss some next. The integration is done over $Z_2$ first conditionally on $Z_1$, so that $Z_2\sim N(\rho Z_1,1-\rho^2)$, and then $Z_1$ is integrated out. Some helpful identities used in the calculations are given in the Appendix. Furthermore, the summations can be evaluated by recursion in $\pd$. In detail, writing $\Theta_\pd=3 \sum_{k=1}^{\pd-1} (\pd-k) H_k$ and $S_\pd = \sum_{k=1}^{\pd-1} H_k$, we initialise $S_1=0$, $\Theta_1=0$, and for $\pd\ge 2$ we have $S_\pd=S_{\pd-1}+H_{\pd-1}$ and $\Theta_\pd=\Theta_{\pd-1}+3S_\pd$.

Remember that in obtaining the skewness we need to divide the third moment by the $\frac{3}{2}$ power of the second, so it is convenient to deal with functions normalised to $\bigmom{\psi(Z)^2}=1$.

\subsection{Examples}

We discuss three functions, drawn in Figure~\ref{fig:n1}.

\begin{figure}[h!]
\begin{center}\begin{tabular}{c}
(a) \scalebox{\figscal}{\includegraphics*{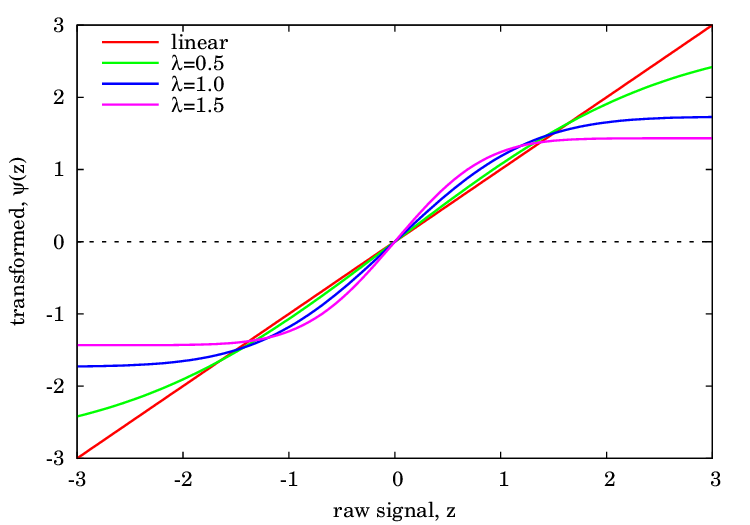}} \\
(b) \scalebox{\figscal}{\includegraphics*{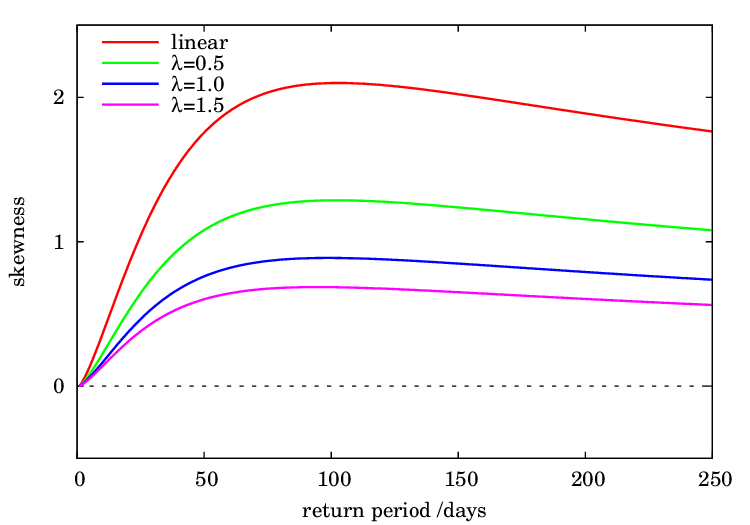}}
\end{tabular}\end{center}
\end{figure}

\begin{figure}[h!]
\begin{center}\begin{tabular}{c}
(c) \scalebox{\figscal}{\includegraphics*{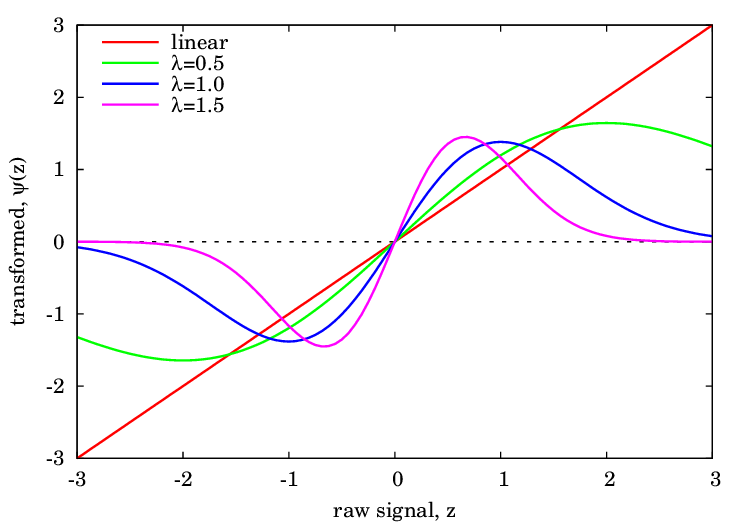}} \\
(d) \scalebox{\figscal}{\includegraphics*{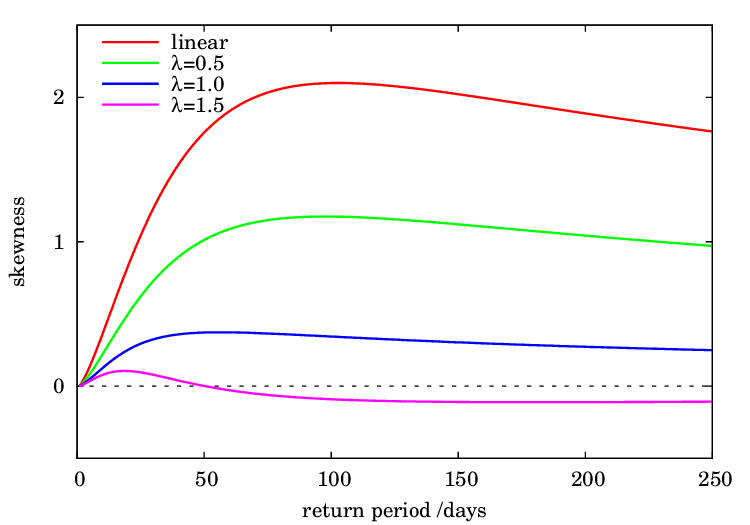}} 
\end{tabular}\end{center}
\end{figure}

\begin{figure}[h!]
\begin{center}\begin{tabular}{c}
(e) \scalebox{\figscal}{\includegraphics*{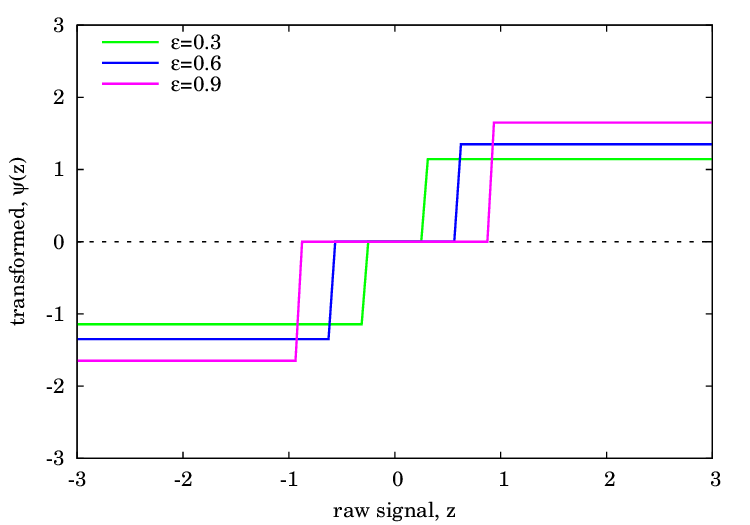}} \\
(f) \scalebox{\figscal}{\includegraphics*{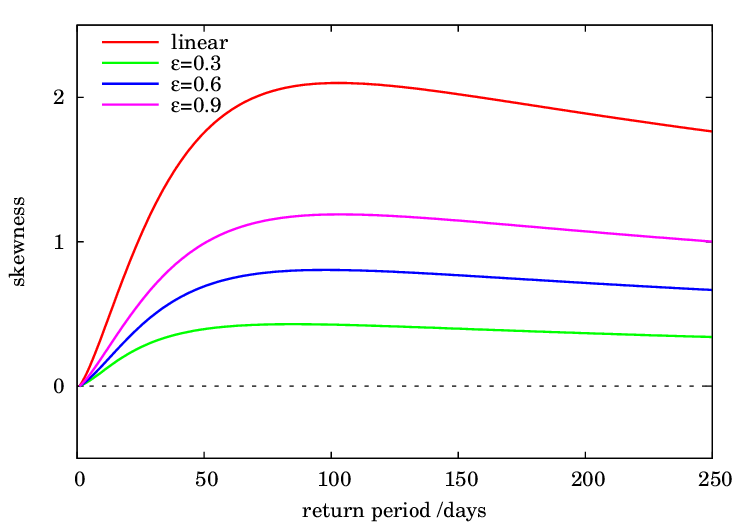}} 
\end{tabular}\end{center}
\caption{Activation functions and their associated term structure of skewness: (a,b) sigmoids with $\lambda=1.0,2.0$; (c,d) reverting sigmoids with $\lambda=0.5,1.0,1.5$; (e,f) double-step with $\varepsilon=0.3,0.6,0.9$.   Using EMA2 type momentum signal throughout, with $N=20,40$.}
\label{fig:n1}
\end{figure}

\clearpage

\subsubsection{Simple sigmoid, $\psi(z)=c_\lambda\cdot\big(2\Phi(\lambda z)-1\big)$}

In effect, this caps the position when the magnitude of the the momentum signal is large. We have
\begin{equation}
H_k = 2a_{k-1}c_\lambda^3  \!\left.
\frac{(2/\pi)^{3/2}\lambda}{\sqrt{1+\lambda^2}} \, \mathrm{arctan} \!\left(\frac{\lambda^2\rho\big/\sqrt{1+\lambda^2}}{\sqrt{1+3\lambda^2+2(1-\rho^2)\lambda^4}} \right)   
\right|_{\rho=R^a_k}
\label{eq:Hk_ss}
\end{equation}
and for normalisation we require $c_\lambda = \big( \frac{2}{\pi}\,\mathrm{arctan}\frac{\lambda^2}{\sqrt{1+2\lambda^2}}\big)^{-1/2}$. We obtain in the limit $\lambda\to0$ that $H_k=2a_{k-1}R^a_k$ which is as expected the linear result.

\subsubsection{Reverting sigmoid, $\psi(z)= c_\lambda\cdot z e^{-\lambda^2 z^2/2}$}

The behaviour of this one is more nonlinear in the sense that it begins \emph{reducing} the position when the momentum gets too high, ultimately to zero if the momentum is strong enough. The rationale for this is that a very strong trend might be more susceptible to reversing (market overbought/oversold), justifying a reduction in position. The maximum positions are held when $z=\pm\lambda\inv$.
We have 
\begin{equation}
 H_k = 2a_{k-1} c_\lambda^3 \!\left.\frac{
\rho \big(1-(1-\rho^2)\lambda^4\big)
}{
\big(1+3\lambda^2+2(1-\rho^2)\lambda^4\big)^{5/2} 
} \right|_{\rho=R^a_k}
\label{eq:Hk_rs}
\end{equation}
and for normalisation we require $c_\lambda= (1+2\lambda^2)^{3/4}$.
Notice that $H_k$ becomes negative if $\lambda$ is high enough: this is not surprising because $\psi'(Z)$ is negative for $|Z|>\lambda\inv$, wherein the model is betting against the trend. As expected $\lambda\to0$ gives back the linear result again.

\subsubsection{Double-step, $\psi(z)=c_\lambda\cdot (\mathbf{1}_{z>\varepsilon}-\mathbf{1}_{z<-\varepsilon})$}

This is $+1$ if the momentum is positive enough, $-1$ if negative enough, and zero at intermediate levels, with the  width of the `dead zone' being $2\varepsilon>0$. This is similar to the one considered by Potters \& Bouchaud. We have\footnote{Do not use these results for $\varepsilon<0$.}
\begin{equation}
H_k = 2a_{k-1}c_\lambda^3 \!\left. 
\phi(\varepsilon) \!\left( \Phi\!\Bigg(\varepsilon\sqrt{\frac{1+\rho}{1-\rho}}\Bigg) - \Phi\!\Bigg(\varepsilon\sqrt{\frac{1-\rho}{1+\rho}}\Bigg) \right)
\right|_{\rho=R^a_k}
\label{eq:Hk_ds}
\end{equation}
and for normalisation we require $c_\lambda= \big(2\Phi(-\varepsilon)\big)^{-1/2}$.

From this it can be seen that as $\varepsilon\to0$, creating a binary response $\psi(z)=\mathrm{sgn}(z)$, the skewness vanishes\footnote{This can also be seen from the sigmoidal case when $\lambda\to\infty$, as the argument in the arctan() term goes to zero. The result can also be seen directly from (\ref{eq:Hkold}), because $\psi(Z_1)^2=1$ a.s.\ and  and $Z_1-\rho Z_2$ is independent of $Z_2$, so the expectation decouples into a product of two expectations each of which is zero. Technical point: The behaviour as $\rho\to\pm1$ is awkward. In fact, this limit is irregular and the result as $\rho\to1$ is not the same as that for $\rho=1$; similarly for $-1$. However, these two values of $\rho$ cannot occur in our problem.}. One way of understanding this is to see that as the position is always of magnitude 1, it does not have the characteristic---associated with positive momentum---of increasing the position when the P\&L is positive.
A less precise explanation is that during periods in which the market is not trending, the strategy loses money rather quickly because it is buying and selling the same size of position as it holds when a trend has been detected. By contrast, the other two (sigmoidal) functions that we have just examined only trade a small size until a trend is established, resulting in P\&L distribution with more, but smaller, losses, and fewer, but bigger, gains: that is where their positive skewness comes from.
A recent piece on FX strategy \cite{Bloom12} makes this point somewhat tangentially. The authors use the simple step function and note that the performance is excellent when the market is trending but very poor otherwise. In the light of what we have just said, we are not surprised by this observation.

As $\varepsilon$ is raised, notice that the skewness rises without limit, which seems rather good: however, if $\varepsilon$ is too high then the algorithm hardly ever trades, so practicalities dictate $\varepsilon\lesssim1.5$.

\begin{figure}[h!]
\centering
\scalebox{\figscal}{\includegraphics*{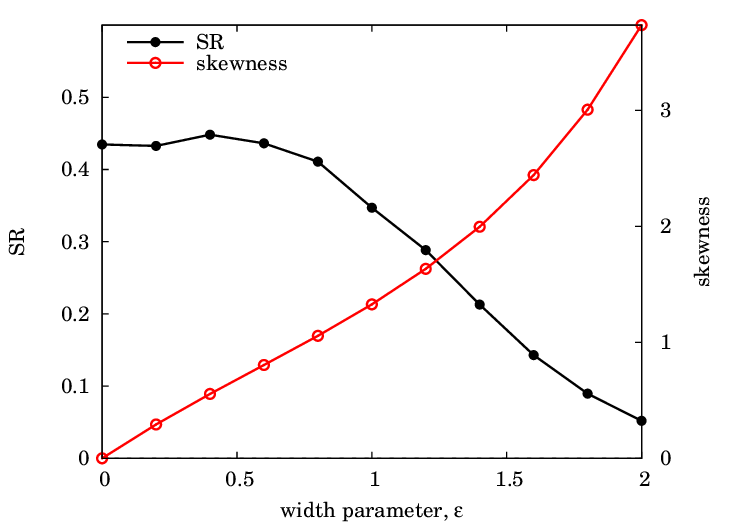}}
\caption{Sharpe ratio and skewness as a function of $\varepsilon$ for double-step type activation function, showing performance dropoff for $\varepsilon>0.6$.}
\label{fig:n2}
\end{figure}

\begin{figure}[h!]
\begin{tabular}{l|rrrrrrr}
$\frac{w_R}{w_S}\setminus\lambda$&	0.25 & 0.50	&0.75	&1.00	&1.25&	1.50&	1.75\\
\hline
$\infty$&\small	0.39\s 1.77	&\small 0.45\s 1.17 &\small 0.48\s 0.69	&\small 0.48\s 0.34	&\small 0.45\s 0.09	&\small 0.40\s $-$0.09 &\small 0.35\s $-$0.23  \\
5.0	&\small 0.38\s 1.74	&\small 0.44\s 1.19 	&\small 0.48\s 0.75	&\small 0.48\s 0.43	&\small 0.46\s 0.21	&\small 0.43\s 0.04 & \small 0.39\s $-$0.09 \\
2.4	&\small 0.38\s 1.73	&\small 0.43\s 1.21	&\small \fbox{0.47\s 0.80}	&\small 0.48\s 0.50	&\small 0.47\s 0.29	&\small 0.45\s 0.14 &\small 0.42\s 0.02 \\
1.5	&\small 0.38\s 1.71	&\small 0.43\s 1.22	&\small 0.46\s 0.83	&\small 0.48\s 0.56	&\small 0.47\s 0.37	&\small 0.46\s 0.22 & \small 0.43\s 0.11 \\
1	&\small 0.38\s 1.70	&\small 0.42\s 1.23	&\small 0.46\s 0.87	&\small 0.47\s 0.62	&\small 0.47\s 0.43	&\small 0.46\s 0.30 & \small 0.45\s 0.19\\
0.67	&\small 0.38\s1.69	&\small 0.41\s 1.24	&\small 0.45\s 0.91	&\small 0.47\s 0.67	&\small 0.47\s 0.50	&\small 0.46\s 0.37 & \small 0.45\s 0.28\\
0.4	&\small 0.37\s1.67	&\small 0.41\s 1.25	&\small 0.44\s 0.94	&\small 0.46\s 0.73	&\small 0.46\s 0.57	&\small 0.46\s 0.46 & \small 0.46\s 0.37\\
0.2	&\small 0.37\s1.65	&\small 0.40\s 1.27	&\small 0.43\s 0.99	&\small 0.45\s 0.80	&\small 0.46\s 0.66	&\small 0.46\s 0.56 & \small 0.46\s 0.48 \\
0	&\small 0.37\s1.63	&\small 0.39\s 1.29	&\small 0.41\s 1.05	&\small 0.43\s 0.89	&\small 0.44\s 0.77	&\small 0.45\s 0.69 & \small 0.45\s 0.62\\
\end{tabular}
\caption{Sharpe ratio and skewness as a function of $\lambda$ and the ratio $w_R/w_S$ for compound sigmoidal activation function.
In each pair of numbers, the first is the SR, and the second is the skewness.}
\label{fig:n3}
\end{figure}

\begin{figure}[h!]
\begin{center}
\scalebox{\figscal}{\includegraphics*{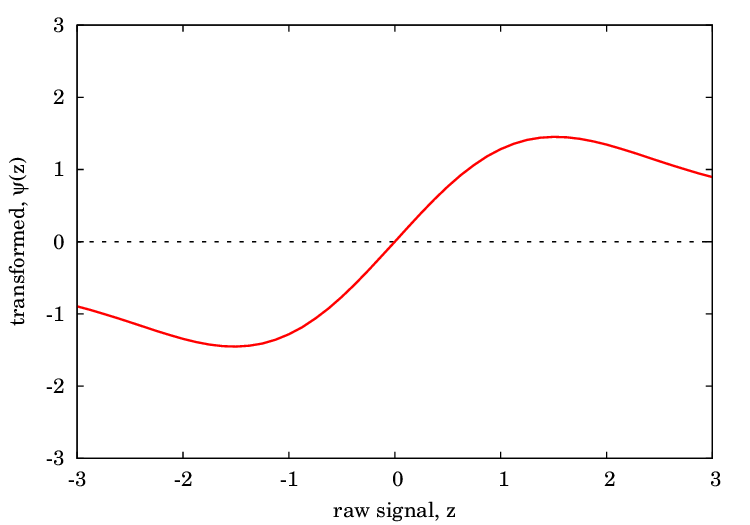}}
\end{center}

\caption{Sketch of the activation function highlighted in Figure~\ref{fig:n3} (parameters: $w_R=0.71$, $w_S=0.30$, $\lambda=0.75$).}
\label{fig:n4}
\end{figure}

\begin{figure}[h!]
\begin{center}
\scalebox{\figscal}{\includegraphics*{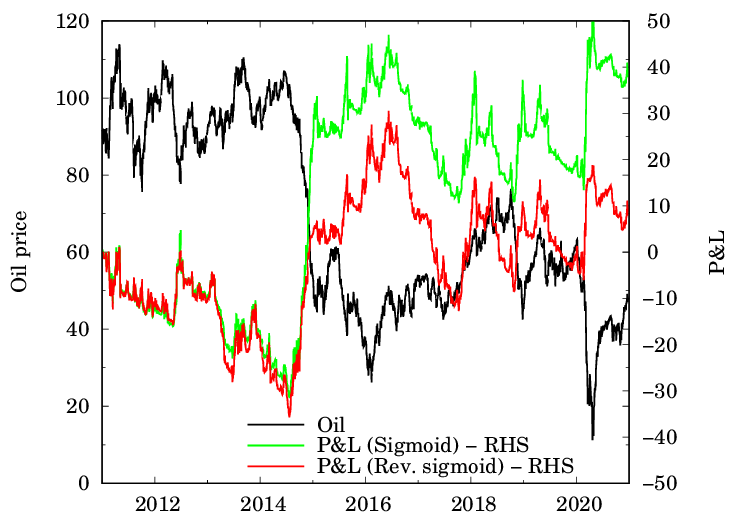}}
\end{center}

\caption{Trend-following oil over a decade, using the sigmoid and reverting sigmoid for activation function ($\lambda=0.75$, $N_1=10$, $N_2=20$). The reverting sigmoid fails to capitalise on the full selloff in late 2014.
}
\label{fig:oil}
\end{figure}

\subsection{Skewness computation}

We show in Figure~\ref{fig:n1} the term structure of skewness for the different examples given above: sigmoid (a,b), reverting sigmoid (c,d), double-step (e,f). The linear result is overlaid for comparison.
The precise choice of momentum crossover does not affect the main conclusion, and we have used EMA2 with $N=20,40$ throughout. Using a faster or slower momentum measure simply stretches or compresses the graph in a horizontal direction, as it did in the linear models.

It is apparent from the results that as the activation function becomes progressively less linear, the main effect is to compress the graph in a vertical direction, so that the maximum skewness is reduced.
With the reverting sigmoid, the graph can be affected much more, to the extent of becoming negative when $\lambda$ is high enough: we predicted this earlier when remarking that $H_k$ could become negative as a result of the activation function being decreasing over much of its domain, so the model spends a lot of time incrementally trading against the trend rather than with it. (In fact for $\psi(z)=ze^{-\lambda^2z^2/2}$ the critical $\lambda$, above which the skewness is no longer everywhere positive, is around 1.3. An explanation is in the Appendix.) 

The general conclusion so far is that any capping effect in the activation function will cause the trading returns to be less positively skewed, and any reverting effect will exacerbate this reduction in skewness. From the perspective of skewness alone, these effects should be avoided as much as possible. However, they may well be justified by reason of risk management and/or expected return, so we consider these next.

\subsection{Empirical analysis and expected return}

Analysis of the expected return is a totally different proposition because there are no theoretical guidelines at all. One can only adopt an empirical approach, seeing what has worked in the past, and relying on it continuing to do so.
\notthis{
For example, if we are preferring the reverting sigmoid to the simple sigmoid (as defined in the previous section), then we are saying that when momentum becomes very strong it is wise to take profits because on average the market tends to reverse. Note however that historical information on this is rare because the momentum factor is only rarely large, so to an extent the design of such strategies is by fiat. The results will necessarily be somewhat subjective because they depend on what data are used.
} 

We need to decide what objective function is to be maximised, and the most natural thing to do is to maximise the Sharpe ratio (SR) of the trading strategy, i.e.\ use an objective function that directly relates to trading model performance. As the SR is the expected return divided by the volatility, we will be penalising any effect that increases volatility without generating enough extra return.
Taking a range of futures contracts across different asset classes (stocks, bonds, FX, commodities) and a range of EMA2 periods (5 vs 10 days, 10 vs 20 days, etc.), we have run trading simulations over the available history, which is typically 20 years or more, and calculated the Sharpe ratio; this gives a list of Sharpe ratios, one for each contract and speed. For simplicity we are going to use the same activation function across all contracts and speeds. We then average the list of Sharpe ratios and use this as our performance indicator, to be maximised\footnote{The degree of temporal and cross-asset-class diversification that can be obtained is governed by what is commonly known as `breadth' and explained in detail by Grinold \cite{Grinold99}.}.

We first examine the double-step activation function. Here there is only one parameter to adjust, namely $\varepsilon$, the half-width of the `dead zone'. Figure~\ref{fig:n2} shows the performance as a function of $\varepsilon$. It is not surprising that the SR drops off as $\varepsilon$ becomes large, because the strategy hardly ever has a position on and can never make any money. What is interesting is that the performance for $\varepsilon<0.6$ is so flat. Thus from the perspective of expected return, one may as well choose any $\varepsilon<0.6$.

However, when we overlay the conclusion about skewness, we can sharpen this deduction. As the skewness has a term structure, we look at one return-period ($\pd$) throughout: we choose $\pd=100$ for convenience, this being the top of the curve for a linear activation function when the EMA periods are 20,40 (see Figure~\ref{fig:n1}(b,d,f)). The skewness is also shown in Figure~\ref{fig:n2}, and clearly it increases with increasing $\varepsilon$, so from that perspective alone we prefer $\varepsilon$ as high as possible. If we can push $\varepsilon$ up to about 0.6 without decreasing the SR, and in doing so can have positive skewness as well, then we should do just that. So this is our first conclusion about design of nonlinear momentum strategies: the blue line in Figure~\ref{fig:n1}(e), $\varepsilon=0.6$, is a good construction.

Next we turn to the sigmoidal functions that we introduced earlier, and take a linear combination of them:
\[
\psi(z) = w_S {\textstyle \big( \frac{2}{\pi}\,\mathrm{arctan}\frac{\lambda^2}{\sqrt{1+2\lambda^2}}\big)^{-1/2}} \big( 2\Phi(\lambda z)-1\big) + w_R (1+2\lambda^2)^{3/4} \, ze^{-\lambda^2 z^2/2} 
\]
with weights $w_{R,S}>0$. To normalise the weighted function we enforce the elliptical constraint
\[
w_S^2 + 2w_Sw_R \cos\delta +  w_R^2=1,
\]
where $\cos\delta$, the correlation between the $R$- and $S$- signals, is given by
\[
\cos \delta =
 \frac{\lambda (1+2\lambda^2)^{1/4}}{1+\lambda^2} \left( \mathrm{arctan}\frac{\lambda^2}{\sqrt{1+2\lambda^2}}\right)^{-1/2} .
\]
The effective number of parameters is now two: the horizontal scaling $\lambda$ and the ratio $w_R/w_S$. The results are shown numerically in the table of Figure~\ref{fig:n3}.

The general picture is that the performance surface is rather flat. Provided one avoids the far left (where the function is too linear and suffers from putting on too much risk when momentum is high) or the top right (where it reverts to zero too quickly when momentum is high), any of the pairs $(\frac{w_R}{w_S},\lambda)$ would do reasonably well.

Again, we overlay the conclusions about skewness, simplifying as before by using the skewness of $\pd=100$-day returns. We know that higher skewness arises from a low value of $\lambda$ and from $w_R/w_S$ small i.e.\ little reversive behaviour. This means going as far as possible to the left of the table, and steering well clear of the top right. Going to the left does lose performance (SR), so some trade-off is required. One particular example is highlighted ($w_R=0.71$, $w_S=0.30$, $\lambda=0.75$) and sketched in Figure~\ref{fig:n4}. It is seen that this does not yield the highest SR, but it is close to the maximum. It does not revert as strongly as the pure $z e^{-\lambda^2 z^2/2}$ does (top row of table), and is likely to be preferable on account of its better third-moment characteristics.

We now return to the discussion at the outset about designing strategies that perform well in specific scenarios. Figure~\ref{fig:oil} shows the results for the sigmoid and reverting sigmoid, with $\lambda=0.75$, $N_1=10$, $N_2=20$, over ten years, for the oil market. It is clear that the reverting sigmoid does substantially less well. This is because the trend is strong and persists for a long time, so the reverting behaviour of the activation function causes severe underperformance. Yet the third column of Figure~\ref{fig:n3} suggests that the reverting sigmoid (top row) on average performs better than the sigmoid (bottom row).  Part of the selling-point of CTA strategies is their ability to produce `alpha' in scenarios such as the selloff in oil (and other commodities, and associated equities) in late 2014. It follows that one should make sure that the strategy does well in such scenarios, rather than simply relying on what has produced the best historical SR. 
This example also corroborates our earlier remark about calibration being sensitive to data history and therefore subjective. Were the oil selloff in late 2014 absent from the calibration, one would arrive at different conclusions about the optimal model.

As a final comment, we see that the smoother activation functions offer only a small improvement in Sharpe ratio over the double-step. However, we have not considered transaction costs, and models that generate sudden large trades can be more difficult to run a large amount of money on. Models that take positions more gradually are therefore easier to handle in trading. They are also easier to handle in backtesting, because a slight change in the definition of the momentum oscillator, which is the input to the activation function, can for a discontinuous function make a huge difference to the simulated position: even minor changes to the strategy can produce unpredictable results.

That said, the characteristic of the double-step function, that it waits until the momentum is above a certain level before trading, may be worthy of further investigation. Thus one aims for a function that is zero for $0<|x|<\varepsilon$, then rises smoothly until a maximum is reached, and then rolls off slowly and asymptotes to a level above zero.

\clearpage

\section{Conclusions and final remarks}

We have shown how to analyse the behaviour of a variety of trend-following models by particular reference to the skewness of the distribution of trading returns. To do this we have needed only the first three moments of the market returns, thus keeping the modelling quite general. As regards linear models the most important formulae are (\ref{eq:mom2lin},\ref{eq:mom3pol}) giving the second and third moments of the trading returns in an elegant application of residue calculus.
 Pure momentum (trending) strategies generate positive skewness even though the market returns might be totally symmetrical. The skewness depends on the return period and has a characteristic term structure which we have derived, illustrated and verified with real data. Hybrid strategies, with trending and counter-trending behaviour, may exhibit a more complex term structure of skewness, and we have shown how to analyse a general linear model.

We have investigated `nonlinear momentum strategies' from different angles, understanding the Sharpe ratio and the skewness---in essence, the first and third moments---of their trading returns. The former was investigated empirically, and the latter mathematically. We have also pointed out that it may be wise to consider the behaviour of a strategy in specific scenarios, especially if they are a \emph{raison d'\^etre} of momentum trading.
Specific conclusions about optimal design are given in the text, but two salient ones are repeated here.

First, the common practice of forming a momentum signal from moving averages and then making a `binary bet' on it, $+1$ or $-1$ according as momentum is positive or negative, is not a good construction;  we make other comments about discontinuous models in the next paragraph. It can be improved by waiting until the signal reaches a threshold before trading. Secondly, although it is a good idea to reduce---not just cap---the position when the momentum is very high, it is not advisable to reduce it too much. If too aggressive, the reducing effect can undesirably cause negatively-skewed trading returns, and much extra trading: when the momentum finally begins to fall, one ends up buying back the position and then dumping it again as the momentum goes back to zero. Also, the reducing effect loses profit badly when the trend is large and prolonged (Figure~\ref{fig:oil}). 

One area that we have not touched upon is the design and analysis of the opposite type of strategy, i.e.\ mean reversion. We can obtain a reversion model from a momentum model simply by multiplying by $-1$. This of course makes the skewness negative, but we do not want it to be as negative as possible. Therefore the optimal design will not simply be the reverse of what we have done here. Instead, something like the reverting sigmoid, now written as $\psi(x) = -c_\lambda x e^{-\lambda^2 x^2/2}$, is likely to be a good idea. When the market deviates from the reversion level a long way, some risk is taken off, which is likely to be beneficial.

A common explanation for the positive skewness is that it arises from the strategy having positive convexity, as mentioned for example in \cite{Dao16}. This is partly true, and we have explained its origin in \S\ref{sec:option}, but in fact there is more to it than that, as will be discussed in forthcoming work.

\clearpage

\appendix

\section{Momentum, price and return (discrete time)}
\label{sec:MPR}

We establish the link between momentum crossovers of prices (total returns), and a weighted sum of returns. The continuous-time analogue is very simple and has been dealt with in eq.(\ref{eq:ctmom1}) et seq.

An elementary view of the simple moving average (EMA1) is that is it the difference between the current price $X_n$ and a weighted average of previous prices $X_{n-j}$. 
Define the average as
\[
\mathcal{E}[X]_n =(1-\alpha) \sum_{j=0}^\infty \alpha^j X_{n-j}
\]
which obeys the update recurrence
\[
\mathcal{E}[X]_n = \alpha \mathcal{E}[X]_{n-1} + (1-\alpha) X_n.
\]
The EMA1 is 
\begin{eqnarray}
X_n - \mathcal{E}_\alpha[X]_n &=& X_n + \sum_{j=0}^\infty \alpha^{j+1} X_{n-j} - \sum_{j=0}^\infty \alpha^j X_{n-j} \nonumber \\
&=& \sum_{j=0}^\infty \alpha^{j+1} (X_{n-j} - X_{n-1-j}) ,
\end{eqnarray}
a weighted sum of returns with weight $\alpha^{j+1}$. Note there is no $(1-\alpha)$ in it and the sum of the weights is not unity.
It admits the simple update formula
\begin{equation}
X_n - \mathcal{E}_\alpha[X]_n = \alpha(X_{n-1} - \mathcal{E}_\alpha[X]_{n-1} \big) + \alpha(X_n-X_{n-1}) ,
\end{equation}
and hence is ``$\alpha\times$ previous value plus $\alpha\times$ most recent return''.

An EMA2 is the difference of two of these. In price terms it is usually formulated as ``fast MA minus slow MA'', which gives
\begin{eqnarray}
\mathcal{E}_\alpha[X]_n - \mathcal{E}_\beta[X]_n 
&=& \big(X_n - \mathcal{E}_\beta[X]\big) - \big(X_n - \mathcal{E}_\alpha[X]_n \big)  \nonumber \\
&=& \sum_{j=0}^\infty (\beta^{j+1}-\alpha^{j+1}) (X_{n-j}-X_{n-j-1}).
\end{eqnarray}
In context $\alpha<\beta$; equivalently $\alpha=1-N_\alpha\inv$ and $\beta=1-N_\beta\inv$ and $N_\alpha<N_\beta$.

When normalising a weighted sum, the premise is that the returns are uncorrelated and of unit variance. If the weights are $(w_j)$ then the variance of the output is simply $\|w\|^2 = \sum_{j=0}^\infty w_j^2$, so one must divide by the square root of that quantity if the variance of the output is to be unity. For EMA1 and EMA2 this is easily obtained by summing the geometric series, and we find respectively that $\|w\|^2= 
\frac{\alpha^2}{1-\alpha^2}$ for EMA1 and $\|w\|^2=\frac{(\alpha-\beta)^2(1+\alpha\beta)}{(1-\alpha^2)(1-\beta^2)(1-\alpha\beta)}$ for EMA2, as used in \S\ref{sec:nonlinear}.

\section{Formulary}

\subsection{Expectation formulae}

If $(Z_1,Z_2)\sim N_2(\rho)$ (the bivariate Normal distribution with $N(0,1)$ marginals and correlation $\rho$) then
\[
\ex\big[ f(Z_1,Z_2) e^{-a_1^2Z_1^2/2} e^{-a_2^2Z_2^2/2} \big] = \frac{1}{\sqrt{D}} \widehat{\ex} \big[ f(\sqrt{D_2/D}Z_1,\sqrt{D_1/D}Z_2) \big]
\]
where under $\widehat{\ex}$, $(Z_1,Z_2)\sim N_2(\widehat{\rho})$, with
\[
\widehat{\rho} = \frac{\rho}{\sqrt{D_1D_2}}, \quad D_i=(1-\rho^2)a_i^2+1, \quad D= (1-\rho^2)a_1^2a_2^2+a_1^2+a_2^2+1
\]

The following results are of use in obtaining (\ref{eq:Hk_ss}--\ref{eq:Hk_ds}).
For $Z\sim N(0,1)$,
\begin{eqnarray*}
\big<Z^{2n}e^{-b^2Z^2/2}\big> &=& (2n-1)!! (1+b^2)^{-(2n+1)/2} \\
\bigmom{\phi(a+bZ)} &=& \frac{1}{\sqrt{1+b^2}} \,\phi\!\left(\frac{a}{\sqrt{1+b^2}}\right) \\
\bigmom{\Phi(a+bZ)} &=& \Phi\!\left(\frac{a}{\sqrt{1+b^2}}\right) \\
\bigmom{Z\phi(a+bZ)} &=& \frac{-ab}{(1+b^2)^{3/2}}\,\phi\!\left(\frac{a}{\sqrt{1+b^2}}\right) \\
\bigmom{Z\Phi(a+bZ)} &=& \frac{b}{\sqrt{1+b^2}}\,\phi\!\left(\frac{a}{\sqrt{1+b^2}}\right) \\
\bigmom{\Phi(aZ)\Phi(bZ)} &=& \frac{1}{2\pi} \mathrm{arctan} \!\left(\frac{ab}{\sqrt{1+a^2+b^2}}\right) + \frac{1}{4}
\end{eqnarray*}
The last result follows from differentiating both sides w.r.t.\ $a$, and using integration by parts on $\bigmom{Z\phi(aZ)\Phi(bZ)}$.

\subsection{Skewness of reverting sigmoid activation function}

We justify why the skewness is always positive for $|\lambda| \lesssim 1.3$.
It was derived for linear models by means of $z$-transforms that for large $\pd$,
\[
\bigmom{ \big(Y^{(\pd)}_n\big)^3 } 
\sim 3\pd \sum_{k=1}^{\infty} H_k.
\]
We are going to calculate this infinite sum, at least approximately, in the EMA1 case. By (\ref{eq:Hk_rs}) the condition for positivity of the above expression is
\[
\sum_{k=1}^\infty 2\alpha^k \frac{\alpha^k\big(1-(1-\alpha^{2k})\lambda^4\big)}{\big( 1+3\lambda^2+2(1-\alpha^{2k})\lambda^4 \big)^{5/2}} > 0.
\]
Write $\alpha^{2k}=u$ and approximate the sum as an integral over $u$, to give
\[
\int_0^1 \frac{u\big(1-(1-u)\lambda^4\big)}{\big( 1+3\lambda^2+2(1-u)\lambda^4 \big)^{5/2}} \frac{du}{u} > 0. 
\]
Upon doing the integral and tidying up, one ends up with
\[
2+9\lambda^2+7\lambda^4-8\lambda^6>0;
\] 
so $\lambda^2< 1.65$.
This does not explain rigorously what goes on in the pre-asymptotic region when $\pd$ is not large, and it uses EMA1 rather than EMA2, but the above analysis seems sufficient.

\clearpage

\bibliographystyle{plain}
\bibliography{}

\end{document}